\documentclass[pra,nofootinbib,twocolumn,amsmath,amssymb,superscriptaddress,a4paper,11pt, accepted=2025-01-10]{quantumarticle}
\pdfoutput=1
\usepackage{graphicx,amsmath,relsize,epstopdf,color,mathtools,bm,braket,rotating}
\usepackage[numbers]{natbib}
\usepackage{url}
\usepackage[hyphenbreaks]{breakurl}
\usepackage{hyperref}
\hypersetup{colorlinks=true,linkcolor=blue,citecolor=blue,urlcolor =blue}
\usepackage[normalem]{ulem}
\usepackage{booktabs}
\usepackage[table,xcdraw]{xcolor}
\usepackage{cancel}

\newcommand{\RE}{\mathop{\mathrm{Re}} \nolimits}
\newcommand{\IM}{\mathop{\mathrm{Im}} \nolimits}
\newcommand{\Tr}{\mathop{\mathrm{Tr}} \nolimits}

\usepackage{soul,xcolor}
\usepackage{graphicx}
\usepackage{dsfont}
\usepackage{placeins}

\begin{document}
	\setstcolor{red}	
	
	\title{Metrological Advantages in Seeded and Lossy Nonlinear Interferometers}
	
	\author{Jasper Kranias}
    \affiliation{National Research Council of Canada, 100 Sussex Drive, Ottawa, Ontario K1N 5A2, Canada}
    \author{Guillaume Thekkadath}
    \affiliation{National Research Council of Canada, 100 Sussex Drive, Ottawa, Ontario K1N 5A2, Canada}
    \author{Khabat Heshami}
    \affiliation{National Research Council of Canada, 100 Sussex Drive, Ottawa, Ontario K1N 5A2, Canada}
    \author{Aaron Z. Goldberg} 
    \affiliation{National Research Council of Canada, 100 Sussex Drive, Ottawa, Ontario K1N 5A2, Canada}

	\begin{abstract}
    The quantum Fisher information (QFI) bounds the sensitivity of a quantum measurement, heralding the conditions for quantum advantages when compared with classical strategies. Here, we calculate analytical expressions for the QFI of nonlinear interferometers under lossy conditions and with coherent-state seeding. We normalize the results based on the number of photons going through the sample that induces a phase shift on the incident quantum state, which eliminates some of the previously declared metrological advantages. We analyze the performance of nonlinear interferometers in a variety of geometries and robustness of the quantum advantage with respect to internal and external loss through direct comparison with a linear interferometer. We find the threshold on the internal loss at which the quantum advantage vanishes, specify when and how much coherent-state seeding optimally counters internal loss, and show that a sufficient amount of squeezing confers to the quantum advantages robustness against external loss and inefficient detection.
	\end{abstract}
	
	\maketitle
\section{Introduction}

An interferometer uses the interference of light passing through two paths to measure the phase difference $\phi$ between them. Interferometers are among the most precise measurement tools in modern science, which is why they are used in many disciplines such as nanoscience \cite{Brueck2005}, optical coherence tomography \cite{Huangetal1991} and its quantum counterpart \cite{Abouraddyetal2002}, magnetometry \cite{BudkerRomalis2007}, and astronomy \cite{Monnier2003}, including the famous gravitational wave detection by LIGO \cite{LIGO2011,Abbottetal2016,LIGO2023squeezing}. 

\begin{figure}[ht]
    \centering
    \includegraphics*[viewport= 10 40 580 770, width=\columnwidth]{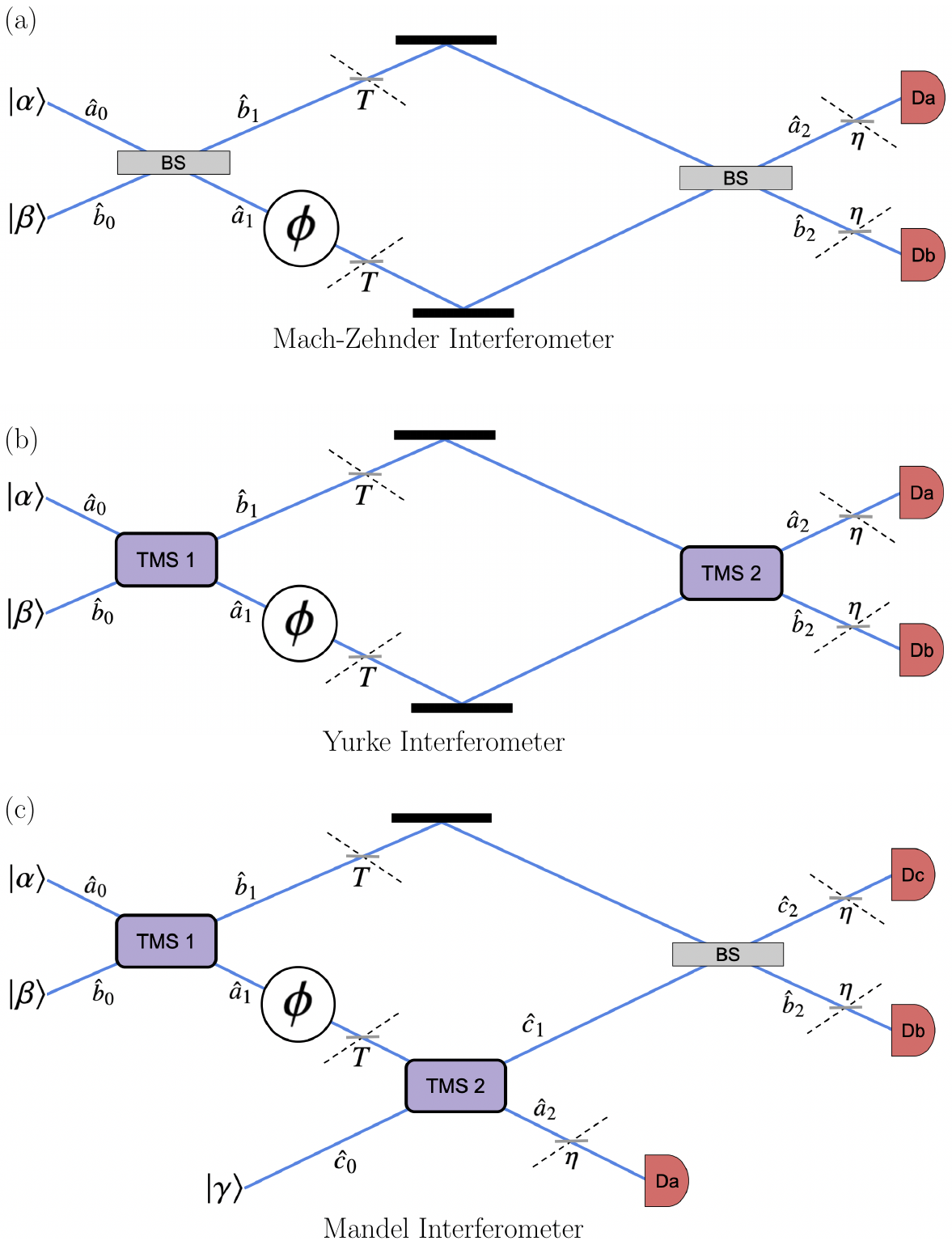}
    \caption{Interferometers for comparison: (a) Mach-Zehnder (linear; SU(2)), (b) Yurke (nonlinear; SU(1,1)), and (c) Mandel (nonlinear), with internal transmission $T$ and external transmission $\eta$. Modes $\hat{a}$, $\hat{b}$, and $\hat{c}$ are seeded with coherent light $|\alpha\rangle$, $|\beta\rangle$, and $|\gamma\rangle$. The light first goes through a 50:50 beam splitter (BS) in (a) or a two-mode squeezer (TMS) in (b) and (c) before the phase shift $\phi$ to be measured is imparted on mode $\hat{a}$. The TMS have squeezing parameters $\zeta_1=r_1$ and $\zeta_2=r_2 e^{i\theta}$ and the light is detected at Da, Db, and Dc.}
    \label{fig:three schemes}
\end{figure}

Reducing the uncertainty of the phase $\phi$ measured by interferometers will improve experimental results and can lead to new applications and discoveries. Interferometry using classical light has its uncertainty bounded by $(\Delta\phi)^2\geq\frac{1}{N}$, where $N$ is the mean number of photons passing through the interferometer. An interferometer using nonclassical states of light can overcome this bound \cite{Caves1981,Xiaoetal1987} to achieve the Heisenberg scaling $(\Delta\phi)^2\geq\frac{1}{N^2}$ \cite{Giovannettietal2004}, but famous nonclassical states such as NOON states \cite{Dowling2008} perform even more poorly than classical light when realistic conditions such as loss are considered \cite{Gilbertetal2008}. Alternatives that avoid fragile probe states have been suggested since the pioneering Yurke interferometer (Fig.~\ref{fig:three schemes}(b)) \cite{SUS}, which uses nonlinear crystals that enact spontaneous parametric down conversion in place of the beam splitters of a classical interferometer. These generally fall into the context of SU(1,1) interferometry, meaning that the interferometer can be viewed as a transformation under the SU(1,1) group of the creation and annihilation operators that represent photons in the interferometer~\cite{SUS,ChekhovaOu2016,Caves2020}, and include Mandel's induced coherence scheme (Fig.~\ref{fig:three schemes}(c)) \cite{Zouetal1991} that has been shown to have some advantages over the Yurke interferometer \cite{Miller2021versatilesuper} and Ref.~\cite{Herzogetal1994}'s scheme that reuses one nonlinear crystal to replace both beam splitters.

Limitations in the power output from nonlinear crystals led to hybrid approaches, whereby seeding down conversion processes (injecting an initial non-vacuum state into one or both of the input modes) with classical states of light can improve the total power while achieving a quantum advantage \cite{InlineSqueezing, Plicketal2010}; this was demonstrated experimentally \cite{Jingetal2011,Hudelistetal2014,Meiretal2023,Gemmelletal2022}. Such seeding has been investigated with the Mandel geometry as well \cite{Cardosoetal2018,Leeetal2019}, and that geometry has been studied extensively \cite{Lemosetal2014,Lahirietal2015,Paterovaetal2018,Vallesetal2018,Lemosetal2023}.
However, total power is not always the best figure of merit: interferometry can measure a phase shift $\phi$ induced when light passes through a sample, but often we are limited by how much light exposure the sample can withstand. Therefore, we always consider the uncertainty that an interferometer can achieve when the amount of light $N_\phi$ passing through the sample is fixed. Even though seeding the Yurke interferometer with coherent light improves phase sensitivity \cite{Florezetal2022arxiv,Gemmelletal2022}, such results do not necessarily mean that seeding is inherently beneficial since seeding increases the amount of light that passes through the sample, and other ways of increasing the amount of light passing through the sample, such as increasing squeezing (the strength of the down conversion process), may provide a greater benefit.

Seeding nonlinear interferometers has been investigated for its advantageous performance in the presence of loss \cite{Ouetal2012,Marinoetal2012,Sparaciarietal2016,Andersonetal2017,Michaeletal2021,Florezetal2022arxiv}, with different results depending upon the location where the loss occurs \cite{Ouetal2012,Marinoetal2012,Michaeletal2021} and the type of detectors used after the interferometer \cite{Andersonetal2017}. In this context, an account of the seeding and squeezing parameters that achieve the optimal phase sensitivity per photon passing through the sample (equivalently, the optimal sensitivity for a fixed number of photons) is incomplete. Here, we analyse the phase sensitivities of seeded Yurke and Mandel interferometers in the presence of loss when $N_\phi$ is fixed.

All of the seeded, lossy, nonlinear interferometers are amenable to the Gaussian state formalism that we briefly outline in Sec.~\ref{GS}. To find the ultimate sensitivities, we utilize the quantum Fisher information (QFI) paradigm and its Gaussian-state formalism due to Ref.~\cite{Safranek} in Sec.~\ref{QFIsec}. We then give descriptions of the different linear and nonlinear interferometers with various geometries in Sec.~\ref{sec:prelim} to set the stage for our results: Sec.~\ref{sec:lossless QFI} finds the conditions for the optimal QFI in terms of squeezing and seeding while taking into consideration the number of photons passing through a sample and Sec.~\ref{sec:lossy qfi} analyzes the same in the presence of loss, comparing each of the interferometers and looking for advantages of the nonlinear ones relative to the linear one. In Sec.~\ref{sec: detection scheme} we provide a detection scheme for the Mandel interferometer using photon number resolving detectors (PNRDs), and find the conditions where it can achieve the minimum uncertainty given by the QFI. No matter the scenario, we show that the QFI only depends on the seeding parameters implicitly via $N_\phi$, such that the QFI can be optimized over all squeezing parameters for a given $N_\phi$ and then we can easily identify the seeding values that contribute the rest of the photons to $N_\phi$. We show that seeding can improve the QFI when there is internal loss (i.e. between the elements of the interferometer); there is a threshold amount of loss beyond which no quantum advantages are possible; and external loss (i.e. right before detection) can be compensated for by sufficiently strong nonlinearities in the Yurke interferometer. The QFI of the Mandel scheme exhibits a striking difference in behaviour when external losses are above or below $50\%$, and its robustness to loss depends on whether mode $\hat{a}$ is measured. The PNRD detection scheme can only surpass the uncertainty of the Mach-Zhender interferometer when external losses are below $50\%$, and only achieves the minimum uncertainty given by the QFI when mode $\hat{a}$ is discarded. Our analytical formulas can be applied to a wide range of interferometry scenarios and help cement the role of seeding in overcoming loss to achieve quantum-enhanced phase sensing.

\section{Methods}
Estimating parameters using quantum states in the presence of loss tends to lead to cumbersome expressions involving sums or integrals that are challenging to optimize. However, for the linear and nonlinear interferometers detailed here, even the lossy quantum states belong to a particular class known as Gaussian states. The latter are particularly amenable to many calculations and, as we presently recount and later develop, have useful properties for quantum parameter estimation.
\subsection{Gaussian States}
\label{GS}
A Gaussian state can be described completely by the mean and covariance of the quadrature operators $\hat{x}_j=\frac{1}{\sqrt{2}}(\hat{a}_j+\hat{a}^\dagger_j)$ and $\hat{p}_j=\frac{1}{\sqrt{2}i}(\hat{a}_j-\hat{a}^\dagger_j)$ in the $j$th mode, where the modes are bosonic, i.e. $\left[\hat{a}_i,\hat{a}_j^\dagger\right]=\delta_{ij}$, and we have employed units where $\hbar=1$. The interferometers we consider are composed only of Gaussian operations, so as long as the initial seeding is Gaussian (we will consider vacuum and coherent state seeding), the resultant state will remain Gaussian as well. In particular, loss can be accounted for within the Gaussian formalism by inserting a fictitious beam splitter and tracing out the fictitious mode.

We write the Gaussian state $\hat{\rho}$ in terms of the displacement vector  $\boldsymbol{d}^{(R)}$ and the covariance matrix $\sigma^{(R)}$, where the label $R$ indicates those quantities to be real; let $\boldsymbol{\hat{d}}^{(R)}=(\hat{x}_1,\hat{p}_1,\ldots,\hat{x}_n,\hat{p}_n)^T$, with the transpose denoted by $^T$, correspond to the components ${d}_i^{(R)}=\Tr\!\left[ \hat{\rho}\hat{d}_i^{(R)} \right]=\langle\hat{d}_i^{(R)}\rangle$, define the centred operators $\Delta\boldsymbol{\hat{d}}^{(R)}=\boldsymbol{\hat{d}}^{(R)}-\langle\boldsymbol{\hat{d}}^{(R)}\rangle$, and the covariance matrix $\sigma_{ij}^{(R)}=\frac{1}{2}\Tr\!\left[ \hat{\rho}\{ \Delta{\hat{d}}_i^{(R)},\Delta{\hat{d}}_j^{(R)}\} \right]=\frac{1}{2}\langle\{{\hat{d}}_i^{(R)},{\hat{d}}_j^{(R)}\}\rangle-\langle{\hat{d}}_i^{(R)}\rangle\langle{\hat{d}}_j^{(R)}\rangle$ (where $\{\cdot,\cdot\}$ denotes the anticommutator).

The initial state of the interferometer is a multimode coherent state $\ket{\alpha_i}$ in each mode. If there are $n$ modes, then the initial state is
\begin{equation}
    \boldsymbol{d}^{(R)}=\sqrt{2}\begin{pmatrix}
        \RE(\alpha_1)\\\IM(\alpha_1)\\ \vdots \\\RE(\alpha_n)\\\IM(\alpha_n)
    \end{pmatrix},\;\;\; \sigma^{(R)}=\frac{1}{2}\mathbb{I}_{2n},
\end{equation}
where $\mathbb{I}_n$ is the $n\times n$ identity matrix. Gaussian unitary operations can be performed by symplectic $2n\times2n$ matrices $F$ that take $\sigma^{(R)}\rightarrow F\sigma^{(R)} F^T$ and $\boldsymbol{d}^{(R)}\rightarrow F\boldsymbol{d}^{(R)}$ while preserving the symplectic form of the group. In general, the operation does not act on all modes of the system. If $f$ is a $2m\times2m$ symplectic matrix which acts on $m$ modes, then $F=P^{-1}\begin{pmatrix}
    f&0\\
    0&\mathbb{I}_{2(n-m)}
\end{pmatrix}P$, where $P$ is a permutation matrix which permutes the modes undergoing the Gaussian operation to be the first modes in the basis. There are also Gaussian operations which displace $\boldsymbol{d}^{(R)}$ by some vector $\boldsymbol{\tilde{d}}$, but we do not use any in the interferometers we consider.

The matrices $f$ for each Gaussian operation we use are given below.

Phase shift by $\phi$:
\begin{equation}
    f_{\mathrm{PS}} \left( \phi \right)=\begin{pmatrix}
        \cos\phi&\sin\phi\\
        -\sin\phi&\cos\phi
    \end{pmatrix}.
\end{equation}
Two-mode squeezing operator $\hat{S}_2(\zeta)=e^{\zeta^*\hat{a}_1\hat{a}_2-\zeta\hat{a}^\dagger_1\hat{a}^\dagger_2}$ with squeeze parameter $\zeta=re^{i\theta}$ (this is an SU(1,1) transformation):
\begin{equation}
\begin{aligned}
    f_{\mathrm{TMS}} \left( r,\theta \right)&=\begin{pmatrix}
        \cosh r\mathbb{I}_2&-\sinh rS_\theta\\
        -\sinh rS_\theta&\cosh r\mathbb{I}_2
    \end{pmatrix},\\
    S_\theta &=\begin{pmatrix}
    \cos\theta&\sin\theta\\
    \sin\theta&-\cos\theta
\end{pmatrix}.
\end{aligned}
\end{equation}
Beam splitter with transmission $\eta$:
\begin{equation}
    f_{\mathrm{BS}}\left(\eta\right)=\begin{pmatrix}
        \sqrt{\eta}\mathbb{I}_2&\sqrt{1-\eta}\mathbb{I}_2\\
        -\sqrt{1-\eta}\mathbb{I}_2&\sqrt{\eta}\mathbb{I}_2
    \end{pmatrix}.
\end{equation}

To model a lossy mode $\hat{a}_i$ with transmission $T$, we add in a fictitious vacuum mode $\hat{a}_f$ by taking 
\begin{equation}
    \boldsymbol{d}^{(R)}\rightarrow\begin{pmatrix}
    \boldsymbol{d}^{(R)}\\0\\0
\end{pmatrix},\;\;\sigma^{(R)}\rightarrow\begin{pmatrix}
    \sigma^{(R)}&0\\
    0&\frac{1}{2}\mathbb{I}_2
\end{pmatrix}
\end{equation}
and performing a beam splitter operation on modes $\hat{a}_i$ and $\hat{a}_f$. Since the fictitious mode is not physically accessible, we trace out mode $\hat{a}_f$ to get the resulting state. The state after the fictitious beam splitter is 
\begin{equation}
    \boldsymbol{d}^{(R)}=\begin{pmatrix}
    \boldsymbol{d}_p\\ \boldsymbol{d}_f
\end{pmatrix},\;\;\sigma^{(R)}=\begin{pmatrix}
    \sigma_{p,p}&\sigma_{p,f}\\
    \sigma_{f,p}&\sigma_{f,f}
\end{pmatrix},
\end{equation}
and we take only the physically accessible part:
\begin{equation}
\boldsymbol{d}^{(R)}\rightarrow\boldsymbol{d}_p,\;\;\sigma^{(R)}\rightarrow\sigma_{p,p}.
\end{equation}

\subsection{Quantum Fisher Information}
\label{QFIsec}
The Quantum Fisher Information (QFI), which we will denote $I_Q$, quantifies the information contained in a quantum state about a parameter $\phi$. The QFI sets a bound on the variance $(\Delta\phi)^2$ in a measurement of $\phi$, which is called the Quantum Cramér–Rao bound. For a single measurement in the asymptotic limit of many repeated measurements \cite{Fisher},
\begin{equation}
\label{QCRB}
    (\Delta\phi)^2\geq\frac{1}{I_Q}.
\end{equation} 
Not every measurement scheme can saturate Eq.~\eqref{QCRB}, but there always exists at least one optimal measurement scheme for which $(\Delta\phi)^2=\frac{1}{I_Q}$. If the state $\ket{\psi}$ undergoes a unitary transformation that is independent of $\phi$, $\ket{\psi}\rightarrow U\ket{\psi}$, then the QFI remains the same since the optimal measurement on $U\ket{\psi}$ is that which takes $U\ket{\psi}\rightarrow U^\dagger U\ket{\psi}=\ket{\psi}$ before performing the optimal measurement on $\ket{\psi}$.

Since the state in the interferometer is Gaussian, it is best to compute the QFI using the Gaussian formalism. Consider a Gaussian state in the complex formalism: $(\sigma,\boldsymbol{d})$ defined by $\boldsymbol{\hat{d}}=(\hat{a}_1,\ldots,\hat{a}_n,\hat{a}^\dagger_1,\ldots,\hat{a}^\dagger_n)^T$ with complex components
${d}_i=\Tr\!\left[ \hat{\rho}\hat{d}_i \right]=\langle{\hat{d}_i}\rangle$, centred complex components $\Delta\boldsymbol{\hat{d}}=\boldsymbol{\hat{d}}-\langle\boldsymbol{\hat{d}}\rangle$, and covariance matrix $\sigma_{ij}=\Tr\!\left[ \hat{\rho}\{ \Delta{\hat{d}}_i,\Delta{\hat{d}}_j\} \right]=\langle\{{\hat{d}}_i,\hat{d}_j\}\rangle-2\langle{\hat{d}}_i\rangle\langle{\hat{d}}_j\rangle$. Note that this is a different convention of writing a Gaussian state than the real form used in \ref{GS}. To convert a Gaussian state from the real convention $(\sigma^{(R)},\boldsymbol{d}^{(R)})$ to the complex form we use
\begin{equation}
\begin{aligned}
\sigma&=2UP\sigma^{(R)} P^TU^\dagger, & \boldsymbol{d}&=UP\boldsymbol{d}^{(R)},
\end{aligned}
\end{equation}
where 
\begin{equation}
    U=\frac{1}{\sqrt{2}}\begin{pmatrix}\mathbb{I}_n&i\mathbb{I}_n\\\mathbb{I}_n&-i\mathbb{I}_n\end{pmatrix}
\end{equation}
and $P$ is the permutation matrix which permutes $(\hat{x}_1,\hat{p}_1,\ldots,\hat{x}_n,\hat{p}_n)^T\rightarrow(\hat{x}_1,\ldots,\hat{x}_n,\hat{p}_1,\ldots,\hat{p}_n)^T$. Note that $\sigma$ is Hermitian.
If the state is pure, then \cite{Safranek}
\begin{equation}
\label{PureGaussQFI}
    I_Q=\frac{1}{4}\Tr\!\left[ \sigma^{-1}(\partial_{\phi}\sigma)\sigma^{-1}(\partial_{\phi}\sigma) \right]+2(\partial_{\phi}\boldsymbol{d})^\dagger\sigma^{-1}(\partial_{\phi}\boldsymbol{d}).
\end{equation}

When there is loss, the interferometer will be in a mixed state. If all modes are in a mixed state, then
\begin{equation}
\label{GaussQFI}
    I_Q=\frac{1}{2}\text{vec}\!\left[ \partial_\phi\sigma \right]^\dagger \mathcal{M}^{-1}\text{vec}\!\left[ \partial_\phi\sigma \right]+2(\partial_\phi\textbf{d})^\dagger\sigma^{-1}(\partial_\phi\textbf{d}),
\end{equation}
where $K=\begin{pmatrix}\mathbb{I}_n&0\\0&-\mathbb{I}_n\end{pmatrix}$, $\mathcal{M}=\overline{\sigma}\otimes\sigma-K\otimes K$, and $\text{vec}\!\left[  \cdot  \right]$ denotes the vectorization of a matrix \cite{Safranek}. For example, if $A=
\begin{pmatrix}a&b\\c&d\end{pmatrix}$, then $\text{vec}\!\left[  A  \right]=\begin{pmatrix}a&c&b&d\end{pmatrix}^T$. 

The ultimate goal, therefore, is to consider various input states, various interferometer geometries, and various loss configurations, all of which leave our states Gaussian, and to determine how to maximize the QFI with respect to the phase shift $\phi$ for each scenario and therefore how to obtain the lowest uncertainty on an estimate of $\phi$ per photon interrogating the phase shift.

\section{Preliminaries}
\label{sec:prelim}
We here perform general computations for the QFI of any Gaussian state that go beyond the previously quoted formulae for $n$-mode and $2$-mode states. We then lay out our linear and nonlinear interferometer setups for analysis.
\subsection{Alternate Form of Lossy QFI}
If there are $n$ modes, computing $\mathcal{M}^{-1}$ directly to find the QFI requires inverting a $4n^2\times 4n^2$ matrix. Therefore, we seek a form of Eq.~\eqref{GaussQFI} that is easier to compute by taking into account the structure of $\mathcal{M}$. If $\sigma K$ has the eigendecomposition $\sigma K=q\lambda q^{-1}$, then
\begin{equation}
\begin{aligned}
\label{altQFI}
    I_Q=\frac{1}{2}\text{vec}\!\left[  \Sigma^T  \right]^T(\lambda\otimes \lambda-\mathbb{I})^{-1}\text{vec}\!\left[  \Sigma  \right]&\\
    +2(\partial_\phi\boldsymbol{d} )^T\tilde{K}q\lambda^{-1}q^{-1}(\partial_\phi\textbf{d})&,
\end{aligned}
\end{equation}
where $\Sigma=q^{-1}(\partial_\phi\sigma)Kq$ and $\tilde{K}=\begin{pmatrix}0&-\mathbb{I}_n\\ \mathbb{I}_n&0 \end{pmatrix}$. A full derivation is given in Appendix~\ref{AA}. 

Instead of inverting a general $4n^2\times 4n^2$ matrix, we have reduced the problem to finding the eigendecompositon of a $2n\times2n$ matrix and inverting a \textit{diagonal} $4n^2\times 4n^2$ matrix, making the computation quicker and easier.

If some modes are pure, then some of the symplectic eigenvalues contained in $\lambda$ will be equal to $1$, making $\lambda\otimes \lambda-\mathbb{I}$ singular \cite{Safranek}. To evaluate the QFI in this case, we can perform a regularization procedure by taking $\sigma\rightarrow\nu\sigma$, which transforms $\lambda\rightarrow\nu\lambda$ and leaves $q$ unchanged. After evaluating Eq.~\eqref{altQFI}, taking the limit $\nu\rightarrow1$ will give the QFI.

\subsection{QFI of Two-Mode Gaussian State}
For a Gaussian state with just two modes, there exists another equation for the QFI. If we define $A=K\sigma$, then the QFI of a two mode Gaussian state is
\begin{widetext}
\begin{equation}
\begin{aligned}
\label{2modeQFI}
    I_Q=\frac{1}{2\left(|A|-1\right)}\Bigg(|A|\Tr\!\left[ (A^{-1}\partial_\phi A)^2 \right] + \sqrt{|I+A^2|}\Tr\!\left[ ((I+A^2)^{-1}\partial_\phi A)^2 \right]&\\
    +4(\lambda_1^2-\lambda_2^2)\bigg(-\frac{(\partial_\phi\lambda_1)^2}{\lambda_1^4-1}+\frac{(\partial_\phi\lambda_2)^2}{\lambda_2^4-1}\bigg)\Bigg) + 2\partial_\phi\boldsymbol{d}^\dagger\sigma^{-1}\partial_\phi\boldsymbol{d}&,
\end{aligned}
\end{equation}
\end{widetext}
where $\lambda_{1,2}=\frac{1}{2}\sqrt{\Tr\!\left[ A^2 \right]\pm\sqrt{(\Tr\!\left[ A^2 \right])^2-16|A|}}$ are the symplectic eigenvalues \cite{Saf2}. This form of the QFI proves to be easier to compute in some cases.

\subsection{Mach-Zehnder Interferometer}
The Mach-Zehnder Interferometer (MZI) (Fig.~\ref{fig:three schemes}(a)) is an SU(2) interferometer that has a phase difference in the paths between two $50:50$ beam splitters \cite{SUS}. If the MZI is seeded with coherent states, then the state will be coherent throughout the entire interferometer, and we can consider it as ``classical" and the paradigm to be linear interferometry.

When seeded with coherent light, the MZI can achieve a phase sensitivity scaling of
\begin{equation}
    (\Delta\phi)^2_{SQL}\sim\frac{1}{N},
\end{equation}
known as the shot noise limit (SNL) or standard quantum limit \cite{Giovannettietal2004}. Here $N$ is the average number of photons which pass through the interferometer (\textit{not} just through the sample).

The diagram in Fig.~\ref{fig:three schemes}(a) is our first depiction of the distinction between internal and external loss; internal loss happens between the first and second beam splitter and external loss before the detectors. When there is no loss, the QFI of the MZI is $I_Q=4N_\phi$. If loss is present, the QFI is that of the lossless case multiplied by the overall transmission; it is multiplied by the internal transmission $T$ and the external transmission $\eta$ to yield
\begin{equation}
    I_Q=4T\eta N_\phi.
\end{equation}
For purposes of comparison, we consider the SNL as $4N_\phi$ since that is the maximum QFI that the coherent seeded MZI can achieve. This is the benchmark against which we compare the other interferometers when seeking quantum advantages.

\subsection{Yurke Interferometer}
The fundamental limit on phase sensitivity is the Heisenberg limit (HL) \cite{Giovannettietal2004}
\begin{equation}
    (\Delta\phi)^2_{HL}\sim\frac{1}{N^2}.
\end{equation}
One way to approach the HL is using SU(1,1) interferometers such as the Yurke Interferometer (Fig.~\ref{fig:three schemes}(b)), which replaces the beam splitters of the MZI with two-mode squeezing \cite{SUS}, where the latter is a nonlinear optical phenomenon. The Yurke interferometer can achieve a sensitivity of $(\Delta\phi)^2=\frac{1}{N(N+2)}$, achieving the Heisenberg scaling $\sim\frac{1}{N^2}$ \cite{SUS}. In Fig.~\ref{fig:three schemes}(b) we again depict the distinction between internal and external loss, as well as allow the interferometer to be seeded by coherent states of light. We will seek to find how to optimally use seeding and squeezing in the Yurke setup to beat the classical limits both in the ideal case and in the presence of loss.

\subsection{Mandel Interferometer}
The Mandel interferometer (Fig.~\ref{fig:three schemes}(c)) uses two-mode squeezing like the Yurke interferometer and so is also a nonlinear interferometer, but instead of performing the second squeezing operation on modes $\hat{a}$ and $\hat{b}$ after the phase shift, a third mode $\hat{c}$ is added and squeezed together with mode $\hat{a}$. Then modes $\hat{b}$ and $\hat{c}$ are mixed on a 50:50 beam splitter before detection. Loss between TMS 2 and the beam splitter is included in the external loss since equal loss in both modes commutes with a beam splitter. Transmission $T$ in the upper arm between TMS 1 and the beam splitter was chosen so that the internal loss is symmetric in a similar manner to Yurke scheme. 

There are a few reasons for considering the Mandel setup instead of the Yurke one. Miller \textit{et al}. argue that the Mandel interferometer has an advantage over the Yurke interferometer when the amount of light that can pass through the sample is limited \cite{Miller2021versatilesuper}. Therefore, we seek to compare it with the Yurke interferometer at fixed $N_\phi$. An additional advantage of the Mandel scheme is that a sample can be probed at frequencies which are not well suited for detection, since high phase sensitivity can be maintained if mode $\hat{a}$ is not detected and one can consider mode $\hat{c}$ to have a different frequency than mode $\hat{a}$ \cite{Miller2021versatilesuper}.

  %\clearpage
%\onecolumngrid
\begin{figure*}[htb]
\centering
%viewport =0 45 600 210
\includegraphics*[width=\textwidth,trim={1cm 1.5cm 0 0},clip]{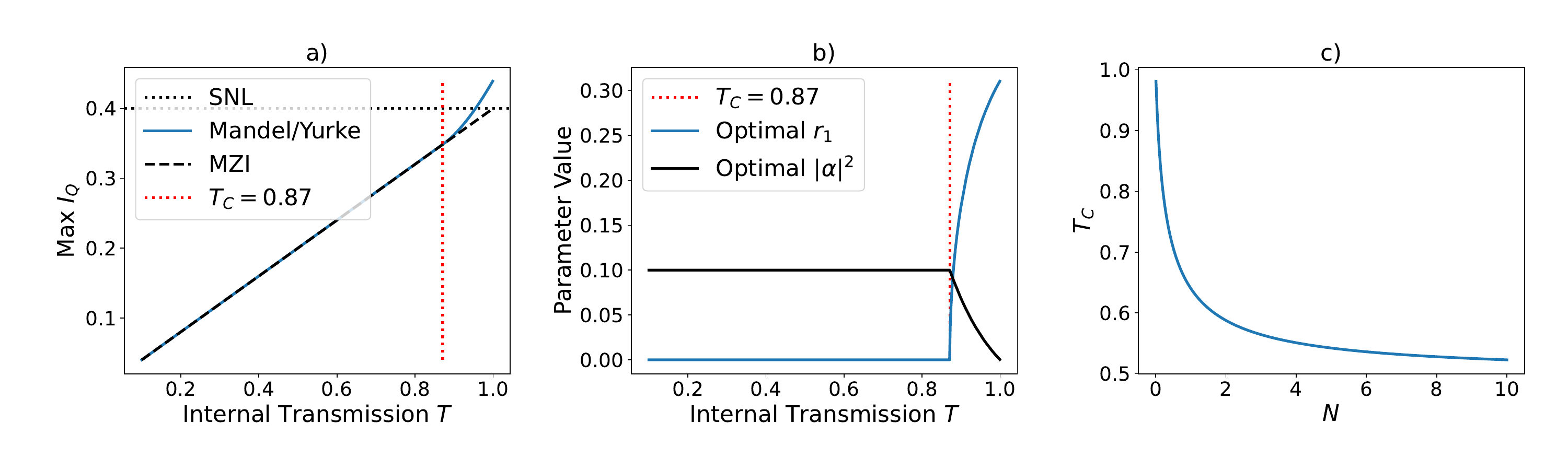}
  \caption{(a) Optimal QFI vs. $T$ when $N_\phi=0.1$ and $\eta=1$. In this case the Mandel and Yurke schemes are equivalent (blue, solid). At $T_C=0.87$ (red, dotted line), they start to surpass the MZI (black, dashed). The Mandel and Yurke schemes can surpass the SNL(grey, dotted), even when there is some internal loss. (b) Squeezing and seeding parameters that optimize the QFI for $N_\phi=0.1$, $\eta=1$ when only mode $\hat{a}$ is seeded with magnitude $|\alpha|^2$. With transmission smaller than $T_C$, it is optimal to have $r_1=0$ (blue line) and use classical light of magnitude $|\alpha|^2$ (black line). For transmission greater than $T_C$ (red dotted line), it becomes beneficial to increase $r_1$ and reduce $|\alpha|^2$. At $T=1$, the lossless case is recovered. (c) Critical transmission $T_C$ vs. $N_\phi$ when $\eta=1$. Below $T_C$ the Mandel/Yurke interferometer has no advantage over the MZI; for larger numbers of photons interrogating the phase shift, more internal loss can be tolerated while retaining a quantum advantage using seeded (when $T<1$) nonlinear interferometers.}
  \label{InternalFig}
  \end{figure*}

  %\twocolumngrid
\section{Results: Lossless Interferometers}
\label{sec:lossless QFI}
How do the nonlinear, seeded interferometers compare to the linear, classical interferometer for phase estimation in the absence of loss?

When there is no loss, the Yurke and Mandel interferometers have the same QFI since they are identical before the phase shift, and all operations after the phase shift are unitary such that an optimal measurement scheme can transform one into the other at will. We here explain how to optimize the amount of information a state can attain about $\phi$ by tuning the input light to probe the sample with the least amount of light $N_\phi$; the full calculations are detailed in Appendix~\ref{AB}.

We calculate the mean number of photons that pass through the sample to be
\begin{equation}
\begin{aligned}
\label{Nphi}
    N_\phi=\cosh ^2r_1|\alpha|^2+\sinh ^2r_1(|\beta|^2+1)&\\
    -2\cosh r_1\sinh r_1\RE(\alpha\beta)&.
\end{aligned}
\end{equation} This depends on both the squeezing and seeding parameters. Note that the total number of photons in the interferometer is \textit{not} twice this number, as asymmetric seeding leads to imbalances between the photon numbers in each arm.

When there are no losses, the QFI of the Mandel and Yurke interferometers can be found from Eq.~\eqref{2modeQFI}:
\begin{equation}
\label{LosslessQFI}
    I_Q=4\cosh^2r_1N_\phi+4\sinh ^2r_1N_\phi-4\sinh ^4r_1.
\end{equation}
For a given number of photons passing through the sample $N_\phi$, we can optimize $I_Q$ over the seeding and squeezing parameter. This results in the maximum
\begin{equation}
\label{IQmax}
    I_{Q_{\mathrm{max}}}=4\cosh^2r_1\sinh ^2r_1=4N_\phi(N_\phi+1)
\end{equation}
when $N_\phi$ is fixed. Equivalently, for a fixed number of photons passing through the sample, the maximum QFI \textit{per photon passing through the sample} is $4(N_\phi+1)$. The maximum is achieved under the condition
\begin{equation}
\label{optcondition}
\sinh^2r_1=N_\phi,
\end{equation}
which together with Eq.~\eqref{Nphi} implies that
\begin{equation}
\label{seedcondition}
\alpha=\sqrt{\frac{N_\phi}{N_\phi+1}}\beta^*.
\end{equation}
Equation~\eqref{seedcondition} can be satisfied by having zero initial seeding ($\alpha=\beta=0$). It is also possible to reach the optimal QFI in the presence of seeding, but seeding does not offer an advantage in the lossless case because the maximum QFI depends only on $r_1$ when $N_\phi$ is fixed. However, if the amount of squeezing $r_1$ is limited by physical constraints such as the nonlinearity of the squeezing device, it is best to make $r_1$ as high as possible, and then seed the inputs with the necessary $\alpha$ and $\beta$ to satisfy Eq.~\eqref{Nphi}. 
Otherwise, improper seeding phases or values can lead to a deterioration in the QFI for a given $N_\phi$: comparing an increase in seeding and squeezing that increase $N_\phi$ by the same amount, the increase in seeding increases the QFI by less than the increase provided by the increase in squeezing.
Although the QFI will no longer be at its maximum, a quantum advantage will remain for any nonzero $r_1$. These together fully inform the use of seeding for attaining quantum-enhanced phase estimation in the lossless scenario for photosensitive samples.

How may the QFI be attained, in practice? Since the states are Gaussian and pure after passing through lossless interferometers, a Gaussian measurement suffices for revealing this maximal information, which is true for any single-parameter measurement on a pure Gaussian state~{\cite{Pineletal2012}}. One such optimal measurement is to run the Gaussian unitaries in reverse after the parameter has been imprinted, followed by homodyne detection~{\cite{Matsubaraetal2019}}. In contrast, lossy interferometers maintain the Gaussianity but not the purity of the state, after which Gaussian measurements are generally no longer optimal (see Ref.~{\cite{Ohetal2019}} for the important proof that Gaussian measurements on Gaussian states are not guaranteed to be optimal measurements). This makes lossy interferometry even more distinct, as we will see in the proceeding section.

\section{Results: Lossy Interferometers}
\label{sec:lossy qfi}
How do the nonlinear, seeded interferometers compare to the linear, classical interferometer for phase estimation in the presence of loss?

When there is loss, the QFIs of the Mandel and Yurke schemes are not necessarily equivalent, since loss is not a unitary operation. Therefore, we compare the lossy QFI of the two schemes to go beyond the equivalence of the previous section and analyze this experimentally relevant scenario of optimal nonlinear interferometry in the presence of loss.

The equations for the QFI in the lossy cases, which will be shown in the next subsections, were found by evolving the state in the Gaussian formalism and using Eq.~\eqref{altQFI} or Eq.~\eqref{2modeQFI}. The QFI equations were each optimized for a fixed $N_\phi$, where the latter is still given by Eq.~\eqref{Nphi} since we only consider loss after the phase shift.\footnote{Loss after the phase shift is the worst-case scenario because internal loss before and after the phase shift have the same effect on the state (loss commutes with phase shifts). For a given total internal loss, any of the loss that occurs before the phase shift will enable the setup to tolerate a higher input power for the same net amount of light going through the sample, thereby increasing the QFI for a given $N_\phi$.} 
As will be seen henceforth, each QFI depends only on the loss parameters $T$ and $\eta$, the number of photons passing through the sample $N_\phi$, and the squeezing parameters $\zeta_1=r_1$ and $\zeta_2=r_2e^{i\theta}$; it only depends on the seeding parameters $\alpha$ and $\beta$ through $N_\phi$ and we have taken $\zeta_1$ to be real without loss of generality.
As such, for a fixed $N_\phi$, the optimal first squeezing parameter will be independent from $\alpha$ and $\beta$ and be given by some $r_1=r_{1\mathrm{opt}} $. In that case, one can find the seeding parameters for achieving the optimal QFI by solving for $\alpha$ and $\beta$ in Eq.~\eqref{Nphi} with $r_1=r_{1\mathrm{opt}} $. While multiple solutions may exist, the optimal QFI can \textit{always} be reached by \textit{only} seeding mode $\hat{a}$, using a coherent state with the magnitude
\begin{equation}
    |\alpha|^2=\frac{N_\phi-\sinh ^2r_{1\mathrm{opt}} }{\cosh ^2r_{1\mathrm{opt}} }.
    \label{eq:seedAopt}
\end{equation} 
It is possible to achieve optimal seeding while seeding both modes, or only mode $\hat{b}$ when $r_{1\mathrm{opt}} \neq0$, but only seeding mode $\hat{a}$ is the simplest method. This allows us to reduce the space of parameters to explore: for a given $N_\phi$, the remaining free parameter to be optimized is only $r_1$ because an optimal seeding arrangement can always be found from it. The optimal seeding is not balanced between the two modes because we explicitly desire the highest QFI \textit{per photon interrogating  the phase shift}, even though the total number of photons in the interferometer is the same regardless of which mode is seeded.
See Appendix~\ref{AC} for further details regarding this calculation.

As before, we report our results in terms of QFI, which provide the ultimate sensitivities possible for any measurement on the states after they pass through the interferometers. It is beyond the scope of our current work to find all of the optimal measurement schemes, so we provide some brief notes here. First, the optimal measurement scheme may be non-Gaussian, as was found in Ref.~{\cite{Ohetal2019}}. Among the Gaussian measurements, homodyne and heterodyne detection may be considered, including schemes with two-mode homodyne or heterodyne detection preceeded by an arbitrary two-mode interferometer. In general, one can specify the most general Gaussian measurements by choosing a complete set of multimode Gaussian states formed from multimode displacements of a fiducial Gaussian state, where the displacements in each mode are along a single quadrature~{\cite{GiedkeCirac2002}}, constructing a positive-operator-valued measure using those states, computing the classical Fisher information from such a measurement, then optimizing the entire procedure over all such complete sets of Gaussian states. This must generally be done separately for each value of the parameters, with different optimal Gaussian measurements found, for example, for each value of loss in the same interferometer, while the actual optimization over fiducial Gaussian states becomes more involved as the number of modes and thus the number of degrees of freedom gets larger.
%This must generally be done numerically for the task at hand when the number of modes and thus number of degrees of freedom in the fiducial Gaussian states gets larger, while particular cases may have analytic solutions that provide complementary insight.

In the subsequent subsections, we consider different locations of where loss can occur in the interferometers and the effects these have on the QFI and the strategies required to optimize the latter. After that we will discuss a particular measurement strategy and investigate how closely it approaches the QFI.

\subsection{Internal Loss}
When only internal loss is considered (ie. $\eta=1$), everything after the loss occurs is unitary, so the Yurke and Mandel interferometers have identical QFIs. The QFI of the Mandel and Yurke interferometers when only internal loss is present is
%\begin{widetext}
\begin{equation}
\label{IQA}
    \begin{aligned}
        I_Q=&\frac{T^2\sinh ^2(2r_1)}{1+2T(1-T)\sinh^2(r_1)}\\
        &+\frac{4T(1+2T\sinh^2(r_1))}{1+4T(1-T)\sinh^2(r_1)}(N_\phi-\sinh ^2r_1).
    \end{aligned}
\end{equation}
%\end{widetext} 
Note, as mentioned above, that this expression depends only on the seeding amplitudes via $N_\phi$. We can optimize this QFI over all $r_1$ for any given $T$ and $N_\phi$ parameters to investigate quantum advantages.
As seen in Fig.~\ref{InternalFig}(a), the Mandel and Yurke interferometers only achieve an advantage over the MZI above a certain transmission threshold $T_C$, which decreases as $N_\phi$ increases (Fig.~\ref{InternalFig}(c)). The exact expression for $T_C$ and arbitrary $N_\phi$ is
\begin{equation}
\label{TCA}
    T_C=\frac{(2N_\phi-1)}{8N_\phi}\left(1\pm\sqrt{1+\frac{16N_\phi}{(2N_\phi-1)^2}}\right),
\end{equation} 
where we take the $+\, (-)$ case for $N_\phi$ above (below) $N_\phi=\frac{1}{2}$, as displayed in Fig.~\ref{InternalFig}(b).
Below this threshold, it is best to seed the interferometer and set $r_1=0$: this is classical light passing through a linear interferometer. Above this threshold, it is best to seed the nonlinear interferometer, with the ratio of light passing through the sample that comes from the seeding versus the squeezing decreasing with increasing transmission. Figures with other values of $N_\phi$ are depicted in Appendix~\ref{app:supp fig}.
Note that $T_C\rightarrow1/2$ as $N_\phi\rightarrow\infty$ and, therefore, that a quantum advantage can never be gained for $T\leq1/2$ no matter how high we make $N_\phi$; this explains the general observation in Ref.~\cite{Andersonetal2017} that the sensitivity resembles classical scalings when losses approach 50\%. 
The optimal seeding $|\alpha|^2$ in mode $\hat{a}$ and squeezing magnitude $r_1$ also depend on $T$, see again Fig.~\ref{InternalFig}(b). 

The overall conclusions for internal loss are thus as follows: in the lossless case, seeding is not beneficial; but, as loss increases, the optimal seeding $|\alpha|^2$ increases and the amount of nonlinearity required $r_1$ decreases; finally, once loss increases such that $T<T_C$, where $T_C$ decreases with $N_\phi$, nonlinear interferometry is no longer superior to linear interferometry with classical light. Therefore, seeding the Mandel or Yurke interferometer is beneficial in the presence of internal loss. The advantages of nonlinear interferometry are maintained for larger internal loss when the sample can tolerate more light going through it but, at the same time, the optimal fraction of light going through the sample that is contributed by nonlinearity decreases and the relative quantum advantage decreases accordingly.

%\begin{figure}[!htb]
%\label{}
%\centering
%\begin{subfigure}{0.4\textwidth}
%    \includegraphics[width=\textwidth]{InternalFigs/II.png}
 %   \caption{}
 %   \label{II}
%\end{subfigure}
%\hfill
%\begin{subfigure}{0.4\textwidth}
 %   \includegraphics[width=\textwidth]{InternalFigs/OI.png}
  %  \caption{}
 %   \label{OI}
%\end{subfigure}
%\hfill   
%\begin{subfigure}{0.4\textwidth}
 %   \includegraphics[width=\textwidth]{InternalFigs/TI.png}
 %   \caption{}
 %   \label{TI}
%\end{subfigure}
%\hfill   
%\caption{a) Optimal QFI vs. $T$ when $N_\phi=0.1$ and $\eta=1$. In this case the Mandel and Yurke schemes are equivalent (blue). At $T_C=0.87$, they start to surpass the MZI (black). b) Parameters which optimize the QFI for $N_\phi=0.1$, $\eta=1$ when only mode $a$ is seeded with magnitude $|\alpha|^2$. Before $T_C$, it is optimal to have $r_1=0$. After $T_C$, it becomes beneficial to increase $r_1$ and reduce $|\alpha|^2$. At $T=1$, the lossless case is recovered. c) Critical transmission $T_C$ vs. $N_\phi$ when $\eta=1$. Below $T_C$ the Mandel/Yurke interferometer has no advantage over the MZI.}
%\label{fig:figures}
%\end{figure}

\subsection{Yurke Interferometer with External Loss}
When there is only external loss (ie. $T=1$), the Yurke and Mandel schemes will each have different a QFI since the non-unitary evolution (loss) occurs after the point at which the schemes become different from each other. In general, the Yurke scheme with $T=1$ has
\begin{widetext}
\begin{equation}
\label{EY}
    \begin{aligned}
        I_Q=\frac{\eta\sinh^2(2r_1)\left(2\eta\gamma+(1-\eta)(\cosh^2(2r_2)-\cos(2\varphi)\sinh^2(2r_2))-(\eta+1)\right)}{2(\gamma-1)(\eta(1-\eta)(\gamma-1)+1)}\\
    +\frac{4\eta\left(\eta\cosh(2r_1)+(1-\eta)\cosh(2r_2)\right)\left(N_\phi-\sinh^2(r_1)\right)}{1+2\eta(1-\eta)(\gamma-1)},
    \end{aligned}
\end{equation}
\end{widetext}
where 
\begin{equation}
    \gamma=\cosh(2r_1)\cosh(2r_2)+\cos\varphi\sinh(2r_1)\sinh(2r_2).
\end{equation}
Unlike the case where there is only internal loss, the QFI depends on $\phi$  and the  relative phase of the two squeeze parameters via $\varphi\equiv\phi+\theta$, but, like the case where there is only internal loss, the QFI only depends on the seeding parameters through $N_\phi$. 

The QFI becomes more robust to external loss as $r_2$ increases and can tolerate over 80\% loss with $r_2\gtrsim3$ (see Fig.~\ref{YurkeExternalFig}(a); the tolerance increases with increasing $N_\phi$ as can be seen in similar Figures in Appendix~\ref{app:supp fig}). Provided that $\sinh^2(r_1)=N_\phi$, when $r_2$ is large the QFI can approach $I_{Q_{\mathrm{max}}}$ [the maximum QFI in the lossless case, given by Eq.~\eqref{IQmax}] within order $\frac{e^{-2r_2}}{\eta}$ at an optimal angle $\varphi_0$ that satisfies 
\begin{equation}
\begin{aligned}
\label{popt}
    \cos\varphi_0&=\frac{-2\sqrt{N_\phi(N_\phi+1)}}{2N_\phi+1}\\
    &\quad+\frac{(1-\eta)\sqrt{N_\phi(N_\phi+1)}}{2(2N_\phi+1)^2\eta\cosh^2(r_2)}+\mathcal{O}\bigg(\frac{e^{-4r_2}}{\eta^2}\bigg).
\end{aligned}
\end{equation} We can observe the $\frac{e^{-2r_2}}{\eta}$ scaling of the distance from $I_{Q_{\mathrm{max}}}$ in Fig.~\ref{YurkeExternalFig}(a); the QFI drops off for low $\eta$ regardless of the value of $r_2$. More specifically, the optimal QFI is
\begin{equation}
    I_Q=I_{Q_{\mathrm{max}}}\left(1-\frac{(1-\eta)(2N_\phi+1)}{2\eta\cosh^2(r_2)}+\mathcal{O}\bigg(\frac{e^{-4r_2}}{\eta^2}\bigg)\right)
\end{equation} when $r_2$ is large.  Note that $\frac{e^{-2r_2}}{\eta}$ is the inverse of the energy reaching the detectors due to TMS 2 alone; large squeezing $r_2$ is required to overcome small external transmission $\eta$. All external loss in the Yurke scheme can, therefore, be compensated for using the second nonlinear element, so long as the phase difference between the pumps of the first and second nonlinear elements are chosen such that it best matches $\phi$.
The optimal angle has a degeneracy of 2, with the first solution $\varphi_0^{(1)}$ being in the range $\!\left[ \frac{\pi}{2},\pi \right]$ and the second solution $\varphi_0^{(2)}=2\pi-\varphi_0^{(1)}$ being in $\!\left[ \pi,\frac{3\pi}{2} \right]$. Fig.~\ref{YurkeExternalFig}(c) shows how $\varphi_0$ changes with $N_\phi$; in the large-photon-number regime, it is best to match the  relative phase of the two nonlinear processes to $\theta=\pi-\phi$. As with the lossless case, the condition $\sinh^2r_1=N_\phi$ requires that Eq.~\eqref{seedcondition} be satisfied by the seeding parameters, which is done most simply by leaving the interferometer unseeded ($\alpha=\beta=0$); initial seeding is less optimal than initial squeezing in the limit of large second squeezing.

%\onecolumngrid
\begin{figure*}[hbt!]
%viewport =0 40 600 210,
\includegraphics*[width=\textwidth,trim={1cm 1.5cm 0 0},clip]{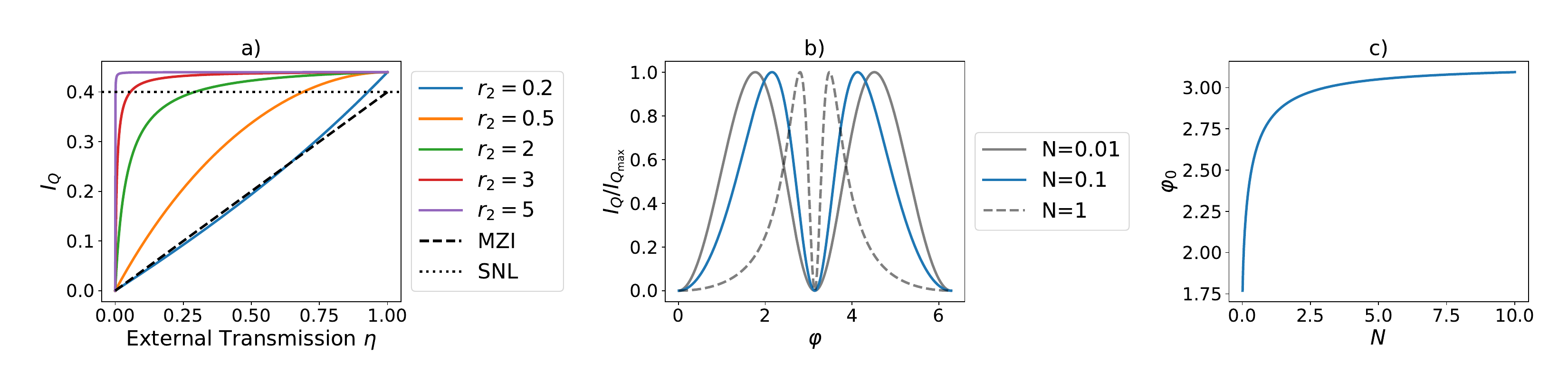}
  \caption{(a) Optimal QFI vs. $\eta$ of the unseeded Yurke interferometer when $N_\phi=0.1$, and $T=1$. When $r_2$ is larger, the QFI becomes more robust to external loss. (b) $\frac{I_Q}{I_{Q_{\mathrm{max}}}}$ vs. $\varphi$ of the unseeded Yurke interferometer with high $r_2$. The QFI is normalized by its maximum when there is no loss, $I_{Q_{\mathrm{max}}}=4N_\phi(N_\phi+1)$, in order to compare the dependence of the curve on $N_\phi$. As $N_\phi$ increases, the peaks become narrower and closer to $\pi$. The QFI is zero at $\varphi=0,\pi,2\pi$ for any finite value of $N_\phi$. This plot was made using $\eta=0.1$ and $r_2=5$, but the curve is almost identical for any value of $\eta$ provided that $r_2$ is sufficiently large. (c) The optimal angle $\varphi_0$ vs. $N_\phi$. This plot shows $\varphi_0^{(1)}$, which lies in $\!\left[ \frac{\pi}{2},\pi \right]$. The other value is $\varphi_0^{(2)}=2\pi-\varphi_0^{(1)}$. As $N_\phi\rightarrow\infty$, both optimal angles will approach $\varphi_0=\pi$.}
  \label{YurkeExternalFig}
\end{figure*}
%\twocolumngrid

We investigate how the QFI changes when $\varphi$ deviates from its optimal value of $\varphi_0$ in Fig.~\ref{YurkeExternalFig}(b). The dependence of $I_Q$ on $\varphi$ is there plotted in the optimal regime of large $r_2$. The larger the value of $N_\phi$, the smaller the window of acceptable values of $\varphi$ before the QFI drops significantly from its maximum. We observe how $\varphi_0$ approaches $\pi$ as $N_\phi$ increases: the two maxima of $I_Q$ versus $\varphi$ are always symmetrically spaced around $\varphi=0$ and this spacing narrows as $N_\phi$ increases, but $I_Q$ vanishes exactly at $\varphi=\pi$ for all finite values of $N_\phi$. Note, then, that the QFI can be very small or even zero when $\varphi$ is not optimal. This speaks to the QFI as governing the asymptotic limit where initial information must be known about $\phi$ should one desire the most precise measurement possible of its value. It also explains how loss can be surmounted: the second nonlinear element acts as a \textit{phase-sensitive} amplifier that is permitted by quantum theory \cite{Cavesetal2012}, unlike noiseless phase-insensitive amplifiers (PIAs) that would lead to deterministic quantum cloning via $\ket{\alpha}\underset{PIA}{\to}\ket{\sqrt{2}\alpha}\underset{BS}{\to} \ket{\alpha}^{\otimes 2}$, and must have its phase tuned such that it amplifies the correct quadrature of the electromagnetic field.

Therefore, the QFI of the Yurke scheme is completely robust to external loss, provided that the second two-mode squeezing magnitude $r_2$ can be made arbitrarily high and $\varphi=\phi+\theta$ can be adjusted to the optimal value. Similar behaviour tolerating up to $80\%$ loss has been demonstrated experimentally in an unseeded Yurke interferometer that has higher gain in the second nonlinear element \cite{ManceauLossTolerant}.

\subsection{Mandel Interferometer with External Loss and Mode $\hat{a}$ Discarded}

Miller \textit{et al}. show that the Mandel scheme can achieve a ``practical advantage'' without detecting mode $\hat{a}$, so that the frequency which passes through the sample can be one which is not well suited for detection, for the lossless scenario \cite{Miller2021versatilesuper}. Therefore we investigate the optimal QFI of the Mandel scheme when mode $\hat{a}$ is discarded in the presence of loss. The QFI of the Mandel interferometer with mode $\hat{a}$ discarded is
%\begin{widetext}
\begin{equation}
\label{MEnoa}
\begin{aligned}
        &I_Q=\frac{\eta\sinh^2(2r_1)\sinh^2(r_2)}{\cosh^2(r_1)\cosh^2(r_2)-1}\\
        &+\frac{4\eta\sinh^2(r_2)(1\!-\eta\!+\!\eta\cosh(2r_1))(N_\phi\!-\!\sinh^2(r_1))}{1+2\eta(\cosh^2(r_1)\cosh^2(r_2)-1)}.
\end{aligned}
\end{equation}
%\end{widetext}
Note that this does not depend on the seeding  of mode $\hat{c}$. The QFI is maximized when  $r_2$ is as large as possible, for two cases depending on how $\eta$ compares to
\begin{equation}
   \eta_0=\frac{1}{2}-\frac{1}{2(N_\phi+1)\cosh^2(r_2)}+\mathcal{O}(e^{-4r_2}).
\end{equation} We consider the large $r_2$ regime where $e^{2r_2}\gg1$. When $\eta>\eta_0$, the maximum QFI is 
\begin{equation}
\label{MandelnoaQFIaboven0}
    I_Q=4\eta N_\phi\bigg(1-\frac{N_\phi}{(N_\phi+1)\cosh^2(r_2)}+\mathcal{O}(e^{-4r_2})\bigg),
\end{equation}
achieved at $r_1=r_{1_{\mathrm{max}}}=\text{arcsinh}(\sqrt{N_\phi})$. This is identical to the QFI of the MZI up to order $e^{-2r_2}$. If $\eta<\eta_0$, the maximum QFI is 
\begin{equation}
\label{MandelnoaQFIbelown0}
    I_Q=2N_\phi\left(1-\frac{1}{2\eta\sinh^2(r_2)}+\mathcal{O}\bigg(\frac{e^{-4r_2}}{\eta^2}\bigg)\right),
\end{equation} achieved when $r_1=0$. This offers loss robustness within order $\frac{e^{-2r_2}}{\eta}$ of $2N_\phi(1-\frac{1}{2\eta_0\sinh^2(r_2)})$, and is completely robust to additional external loss below $\eta=\frac{1}{2}$ provided $r_2$ is large enough (since $\eta_0\to\frac{1}{2}$ for large $r_2$), as seen in Fig.~\ref{MandelnoaFig} and with other values of $N_\phi$ in Appendix~\ref{app:supp fig}. Regardless, the Yurke scheme significantly outperforms the Mandel without mode $\hat{a}$ for all values of $\eta$ because said Mandel scheme cannot achieve Heisenberg scaling, demonstrating the drawback of detecting light at a wavelength different from the one that probes the phase shift. We should expect the Mandel scheme to not achieve Heisenberg scaling for the $\eta<\eta_0$ case since the light probing the phase shift when $r_1=0$ is coherent and, therefore, independent from the other modes. When $\eta>\eta_0$, squeezed light does interrogate the phase shift in the optimal scenario, but information beyond the capabilities of the MZI can not be extracted without detecting mode $\hat{a}$.

Unlike the Yurke scheme with external loss, the QFI here is independent of $\phi$ and $\theta$, so phases do not need to be tuned to achieve optimal QFI. However, since the QFI effectively enacts the optimal measurement scheme, it is entirely possible that the latter depends explicitly on all of these phases.

Miller \textit{et al}. found that, for certain measurement schemes using only modes $\hat{b}$ and $\hat{c}$ and no loss, the Mandel interferometer achieves a phase uncertainty smaller than that of the MZI when they are compared with equivalent total resources \cite{Miller2021versatilesuper}. However, when we compare them using an equivalent number of photons passing through the sample, the Mandel interferometer does not have a \textit{quantum} advantage in the case considered. Moreover, an examination of Eq.~(4) in Ref.~\cite{Miller2021versatilesuper} reveals that it has classical $\sim\frac{1}{N}$ scaling, as expected from our QFI  values in Eqs.~\eqref{MandelnoaQFIaboven0}-\eqref{MandelnoaQFIbelown0}. The actual practical advantage achievable with the Mandel scheme is in the presence of loss: when external loss is greater than 50\%, the Mandel scheme leaving mode $\hat{a}$ undetected can outperform the MZI by an arbitrarily large factor of $1/(2\eta)$ larger QFI. The advantage of the Mandel scheme is to amplify the state prior to photons being lost from it, which could be incorporated into the MZI by similarly amplifying the signal prior to the occasion for external loss to occur.

\begin{figure}[hbt!]
    \includegraphics[width=\columnwidth,trim={1cm 1.4cm 0 0},clip]{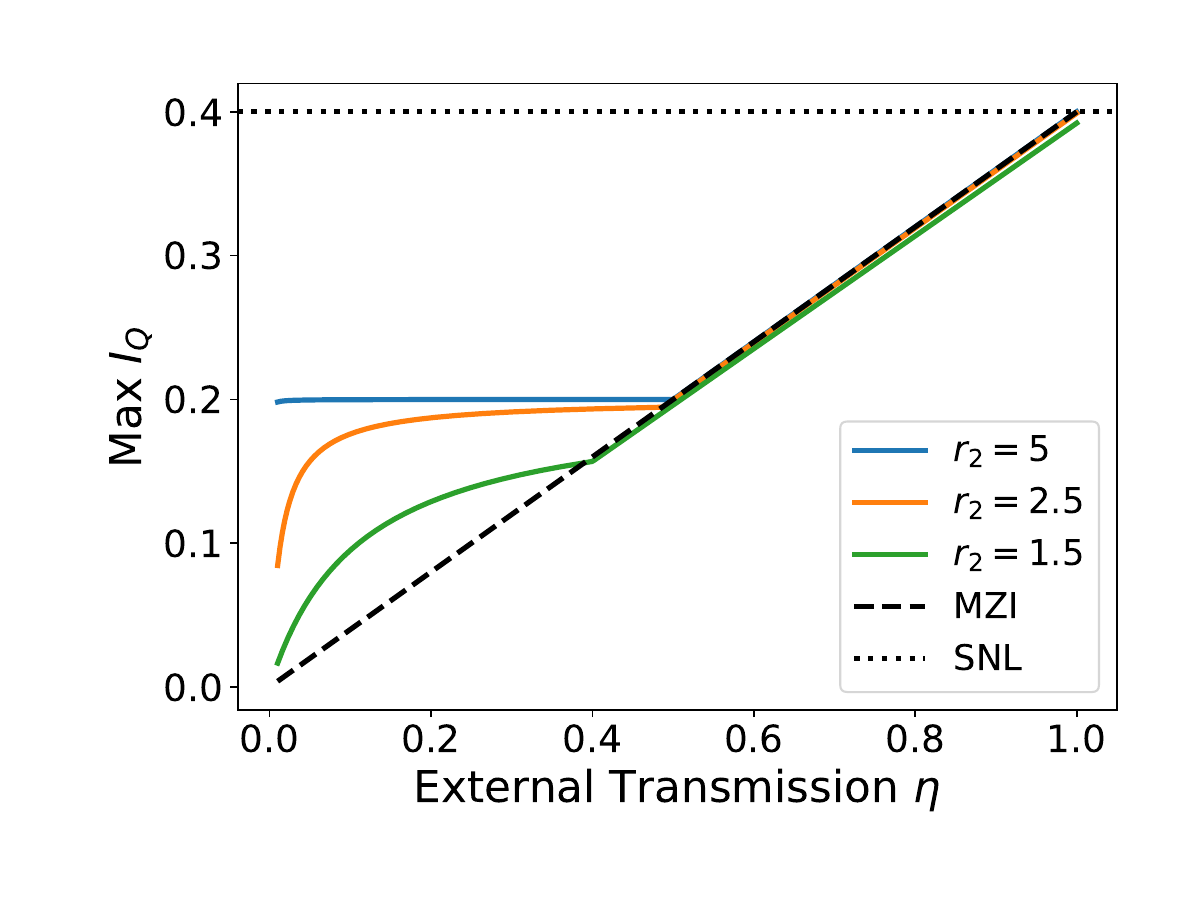}    
\label{MEnoaPic}
\caption{Optimal QFI vs. $\eta$ of the Mandel interferometer when mode $\hat{a}$ is discarded for $N_\phi=0.1$ and $T=1$, for various values of $r_2$. Above $\eta=\eta_0$ there is no advantage over the classical MZI, and we can see that the QFI reduces slightly when $r_2$ is smaller. However, below $\eta=\eta_0$, the scheme becomes robust to  external loss beyond $\eta_0$ up to order $\frac{e^{-2r_2}}{\eta}$. We can see that as $r_2$ increases, $\eta_0$ is larger and the QFI drops off less as $\eta$ gets smaller. }
\label{MandelnoaFig}
\end{figure}

\subsection{Mandel Interferometer with External Loss and All Modes Retained}
The QFI of the Mandel interferometer with external loss when all modes are included is
\begin{widetext}
\begin{equation}
\label{MQFI}
\begin{aligned}
    I_Q= &\frac{\eta\sinh^2(2r_1)(2\eta(1-\eta)(1+\cosh^2(r_1)\cosh^4(r_2))+(2\eta-1)\cosh^2(r_2)(\cosh^2(r_1)-(1-\eta))-1)}{(\cosh^2(r_1)\cosh^2(r_2)-1)(1+2\eta(1-\eta)(\cosh^2(r_1)\cosh^2(r_2)-1))} \\ 
    & \hspace{3cm} +\frac{4\eta(1-\eta+\eta\cosh(2r_1))(\eta+(1-\eta)\cosh(2r_2))(N_\phi-\sinh^2(r_1))}{1+4\eta(1-\eta)(\cosh^2(r_1)\cosh^2(r_2)-1)}.
    \end{aligned}
\end{equation}
\end{widetext}
As when mode $\hat{a}$ is discarded, the QFI does not depend on the seeding of mode $\hat{c}$. When $\eta<\frac{1}{2}$ and we are in a large $r_2$ regime ($e^{2r_2}\gg1$), the maximum QFI is
\begin{equation}
    I_Q=2N_\phi\left(1-\frac{(1-2\eta)}{4\eta(1-\eta)\sinh^2(r_2)}+\mathcal{O}\bigg(\frac{e^{-4r_2}}{\eta^2}\bigg)\right).
\end{equation}
achieved at $r_1=0$. Like when mode $\hat{a}$ is discarded, this shows robustness to external loss up to order $\frac{e^{-2r_2}}{\eta}$, but including mode $\hat{a}$ provides this for $\eta<\frac{1}{2}$ even when $r_2$ is large but finite. However, when $\eta>\frac{1}{2}$, including mode $\hat{a}$  allows us to surpass the MZI. The optimal QFI and corresponding value of $r_2$ depends on the condition 
\begin{equation}
\label{r20cond}
    N_\phi>\frac{2(1-2\eta(1-\eta))+\sqrt{2(1-2\eta(1-\eta))}}{2\eta(2\eta-1)}-1.
\end{equation}
\begin{figure*}[hbt!]
    \includegraphics[width=\textwidth,trim={1cm 1.4cm 0 0},clip]{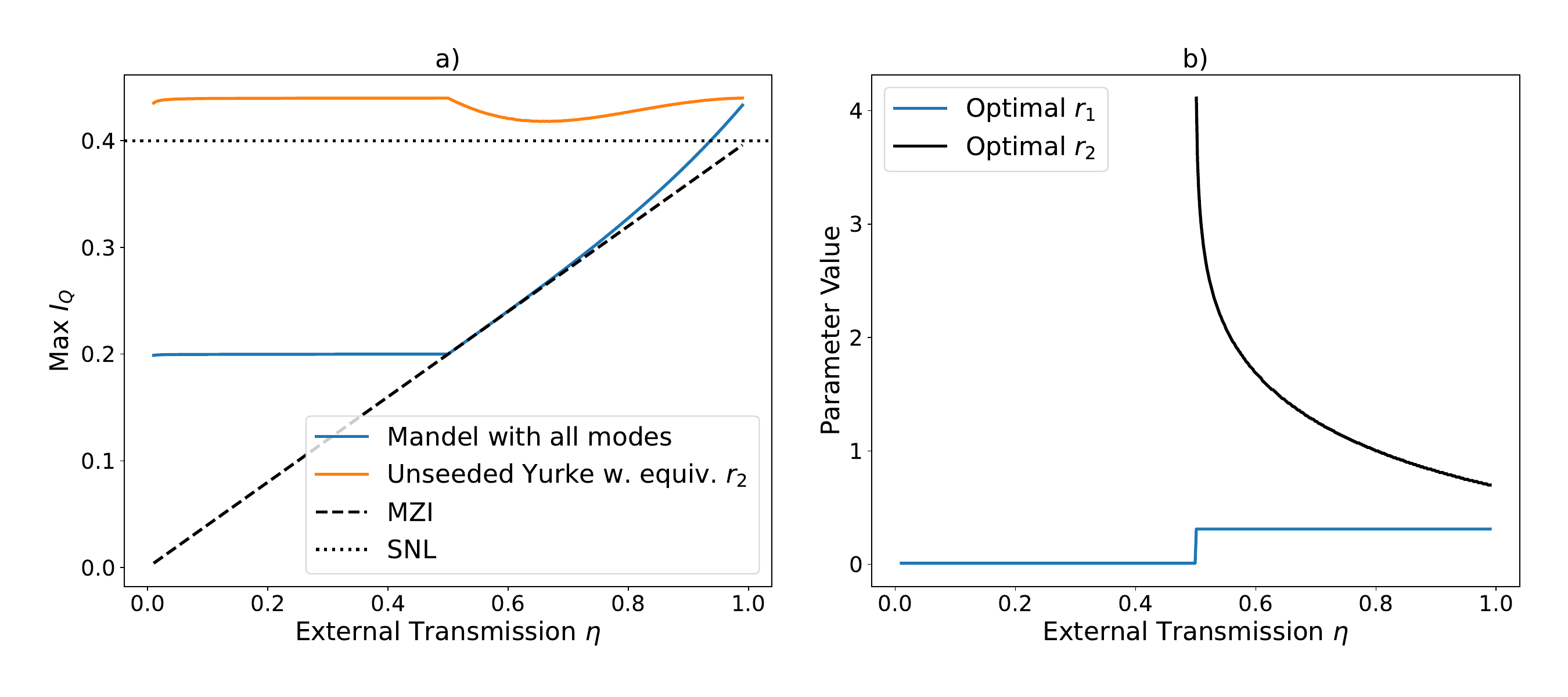}
\caption{(a) Optimal QFI vs. $\eta$ of the Mandel interferometer with all modes considered for measurement, with $N_\phi=0.1$ and $T=1$. The QFI has a quantum advantage which lessens with loss until the QFI is equivalent to the MZI at $\eta=\frac{1}{2}$. After $\eta=\frac{1}{2}$, there is a complete robustness to external loss (provided that $r_2$ is high enough), giving the Mandel scheme a practical advantage over the MZI. When generating this plot numerically the maximum value of $r_2$ was set to $r_2=5$; if $r_2$ has a lower maximum then the QFI will be reduced for low values of $\eta$. To compare the Yurke and Mandel schemes, the QFI of the unseeded Yurke interferometer using the optimal $r_1$ and $r_2$ values of the Mandel interferometer is shown. For all values of $\eta$, the QFI of the Yurke scheme is greater than the QFI of the Mandel scheme and surpasses the SNL, despite the Yurke interferometer using the optimal $r_2$ of the Mandel scheme and not its own. (b) Optimal parameters of the Mandel interferometer with all modes considered for the optimal measurement scheme. When $\eta<\frac{1}{2}$, it is optimal to have $r_1=0$ (no squeezing in the first nonlinear element, only seeding) and $r_2$ as large as possible. When $\eta>\frac{1}{2}$, it is optimal to have $r_1=r_{1_{\mathrm{max}}}$ (no seeding), and $r_2=r_2^{(1)}$ as given in Eq.~\eqref{MEar21}, since Eq.~\eqref{r20cond} is not satisfied for any $\eta$ in the range $\frac{1}{2}<\eta<1$ when $N_\phi=0.1$.}
\label{MandelyesaFig}
\end{figure*}
If Eq.~\eqref{r20cond} is satisfied, then the optimal value of $r_2$ is
\begin{equation}
    r_2^{(0)}=0,
\end{equation}
which results in a QFI of
\begin{equation}
    I_Q=\frac{4\eta^2 N_\phi(N_\phi+1)}{1+2N_\phi\eta(1-\eta)}.
\end{equation}
If Eq.~\eqref{r20cond} is not satisfied, then the optimal value of $r_2$ and the QFI it results in are
\begin{widetext}
\begin{equation}
\label{MEar21}
    r_2^{(1)}=\text{arccosh}\left(\sqrt{\frac{2(1-2\eta(1-\eta))+\sqrt{2(1-2\eta(1-\eta))}}{2\eta(2\eta-1)(1+N_\phi)}}\right),
\end{equation}
\begin{equation}
\begin{aligned}
    I_Q=4\eta N_\phi\Bigg(1+\eta N_\phi\left(1-2(1-\eta)(2\eta-1)-2(1-\eta)\sqrt{2-4\eta(1-\eta)}\right)\Bigg).
\end{aligned}
\end{equation}
\end{widetext}
Note that Eq.~\eqref{r20cond} will be satisfied more easily when both $N_\phi$ and $\eta$ are larger: larger $N_\phi$ tolerates more loss (smaller $\eta$) and larger $\eta$ tolerates weaker probes (smaller $N_\phi$) while still retaining an quantum-enhanced scaling for the QFI. We can interpret this as a limit on amplification increasing the QFI; when $N_\phi$ is large enough, amplification is only necessary beyond a certain amount of loss. As seen in Appendix~\ref{app:supp fig}, when $N_\phi$ becomes larger, the amount of loss needed before amplification becomes beneficial increases. When losses exceed $50\%$, the optimal scenario changes and it is best to have $r_1=0$ and as much amplification as possible.

The optimal QFI surpasses that of the MZI for both $\eta<\frac{1}{2}$ and $\eta>\frac{1}{2}$. As seen in Fig.~\ref{MandelyesaFig}(a), the quantum advantage above $\eta>\frac{1}{2}$ does not display the absolute noise robustness of the Yurke scheme with external loss. Moreover, the Mandel scheme is inferior to the Yurke scheme in terms of robustness to external loss, even when all modes are included. The optimal squeezing magnitudes $r_1$ and $r_2$ are shown in Fig.~\ref{MandelyesaFig}(b) for $N_\phi=0.1$. 

Interestingly, when $\eta<1/2$ the optimal scenario has no seeding and all squeezing at the first nonlinear element, while when $\eta>1/2$ the optimal scenario has no squeezing and all seeding at the first nonlinear element. This is in contrast to the case of internal loss, for which transmission $T<T_C$ (where $T_C>1/2)$ had an optimal scenario of no squeezing and all seeding, displaying the intricate distinctions between internal and external loss.

As with the Mandel scheme when mode $\hat{a}$ is discarded, the QFI does not depend on any phases, but this is not necessarily an advantage over the Yurke scheme since the sensitivity of a particular measurement that attains the QFI can  be phase dependent. As well, even though detecting mode $\hat{a}$ is unnecessary to achieve  loss robustness when $\eta<1/2$, it is possible that the optimal measurement may be more plausible when all modes are considered (this is not a situation we consider strongly because the advantage is explicitly in avoiding measurement of mode $\hat{a}$).
\begin{figure}[htb]
    \includegraphics[width=\columnwidth,trim={1cm 1.4cm 0 0},clip]{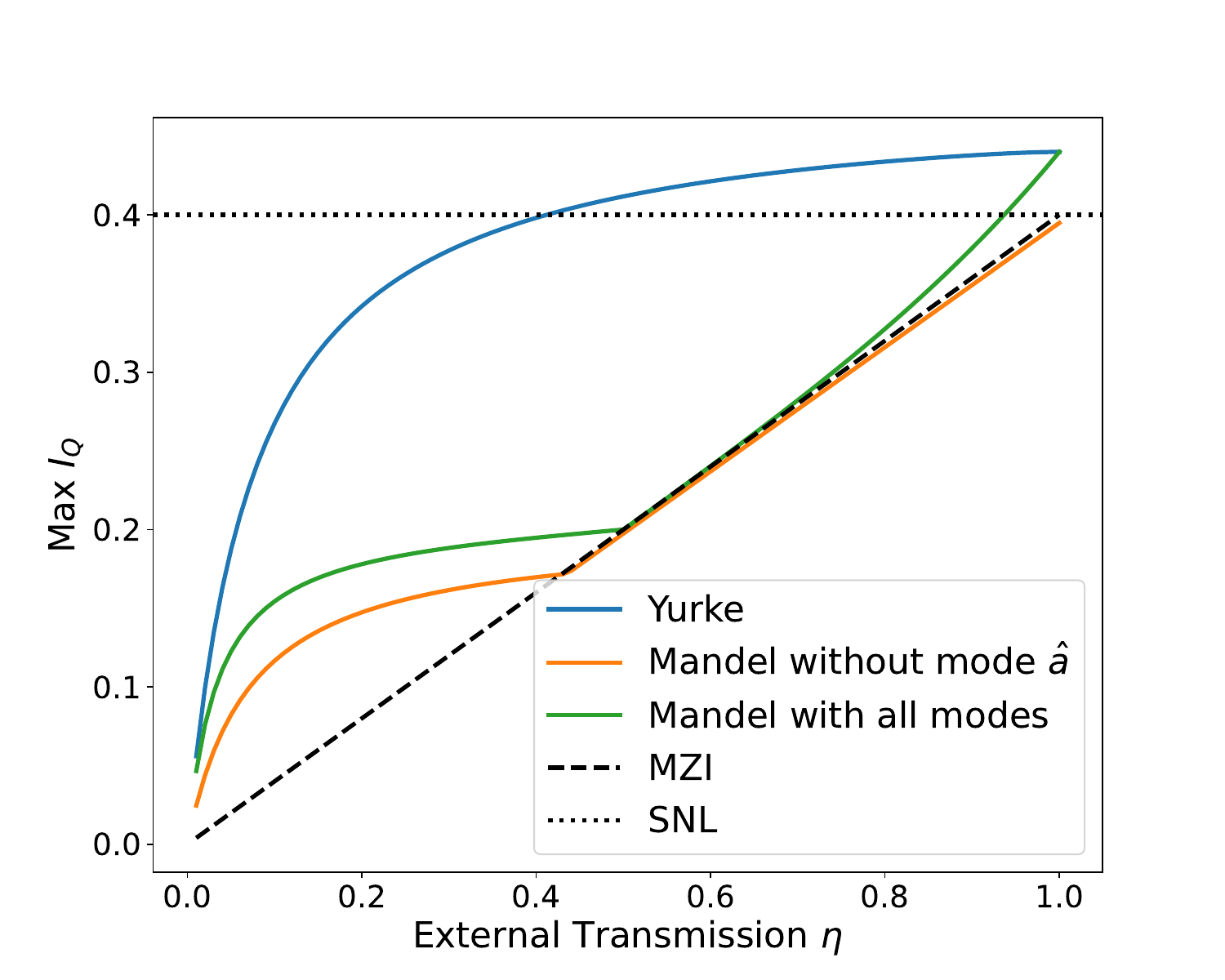}    
\caption{Optimal QFI vs. $\eta$ of the three interferometer schemes for $N_\phi=0.1$, $T=1$, and the experimentally achievable value $r_2=1.7$. The Yurke scheme significantly outperforms the others at all values of $\eta$. We also see that the Mandel scheme is robust to more external loss when mode $\hat{a}$ is included. All schemes scale in powers of $\frac{e^{-2r_2}}{\eta}$ when amplification is used to overcome external loss.}
\label{ComparisonFig}
\end{figure}

When $\eta>\frac{1}{2}$ the optimal value of $r_2$ for the Mandel scheme is finite, which is not the case with the Yurke interferometer. However, as shown in Figure~\ref{MandelyesaFig}(a), the Mandel scheme's favouring of a finite $r_2$ is not an advantage over the Yurke scheme since the Yurke interferometer outperforms the Mandel interferometer when it uses the same finite $r_2$. Similar results for other values of $N_\phi$ are depicted in Appendix~\ref{app:supp fig}. It should be noted that squeezing beyond 10 dB has been achieved by numerous research groups \cite{Schonbecketal2018,Arnbaketal2019,Shietal2020,Meylahnetal2022,Shajilaletal2022}, with records such as Ref.~\cite{Vahlbruchetal2016}'s 15 dB of squeezing corresponding to a maximum squeezing amplitude of $r\approx 1.7$. Fig.~\ref{ComparisonFig} shows a comparison of all three interferometer schemes at $r_2=1.7$ for $N_\phi=0.1$, and similar plots for other values of $N_\phi$ are shown in Appendix~\ref{app:supp fig}.

Miller \textit{et al}. also point out that the Mandel scheme is advantageous over the Yurke interferometer when the detectors are sensitive to light \cite{Miller2021versatilesuper}. The Mandel interferometer may prove advantageous over the Yurke interferometer if we normalize by the amount of light which reaches the detectors instead of that which passes through the sample.

\subsection{Equal Squeezing}

\begin{figure}[hbt!]
    \includegraphics[width=\columnwidth,trim={1cm 1.4cm 0 0},clip]{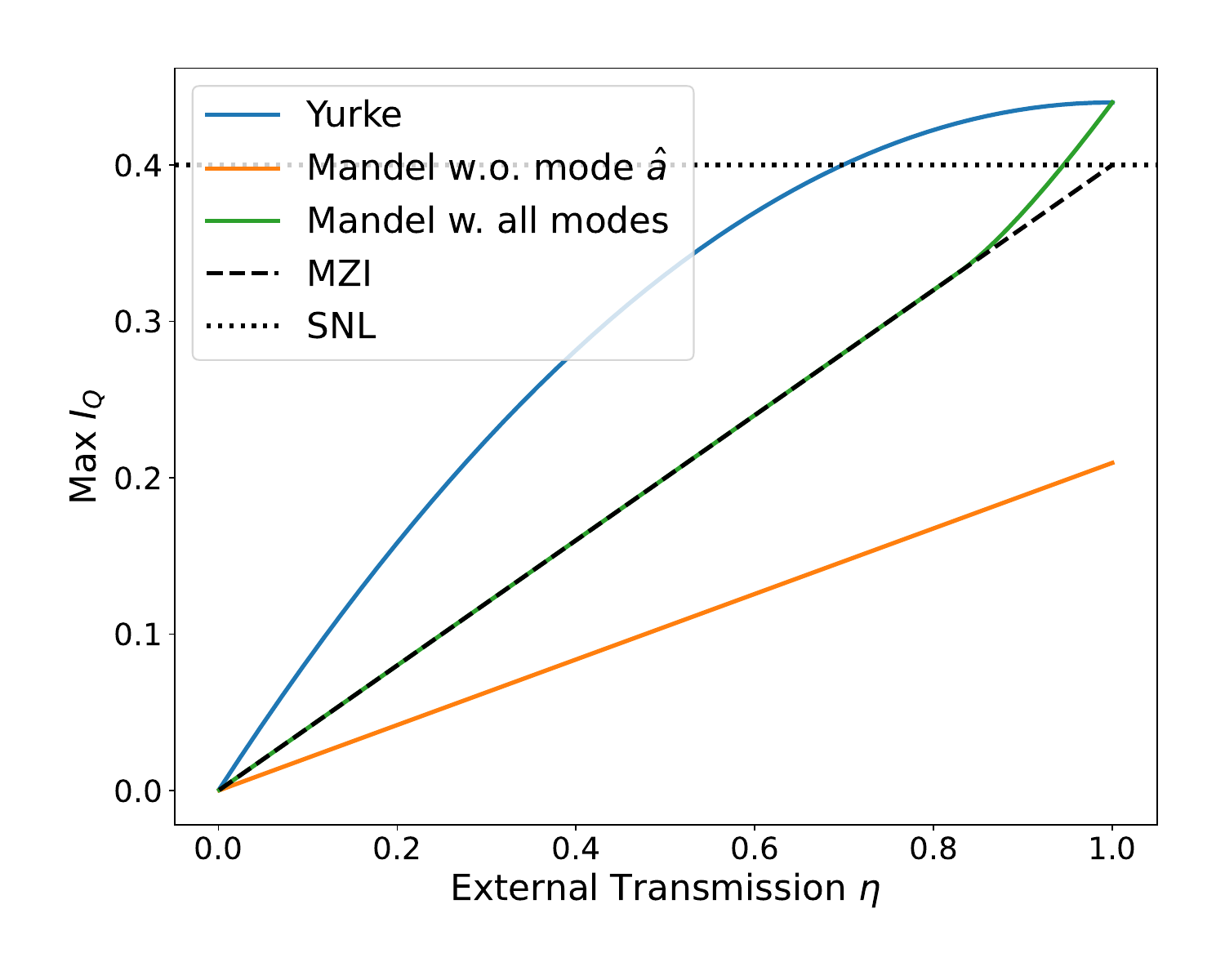}    
\caption{Optimal QFI vs. $\eta$ of the three interferometer schemes for $N_\phi=0.1$ and $T=1$ when $r_1=r_2$. The Yurke scheme significantly outperforms the others at all values of $\eta$. Including mode $\hat{a}$ allows the Mandel scheme to go beyond linear scaling with $\eta$ and surpass the SNL. Without mode $\hat{a}$ the Mandel scheme is not only always linear, but has smaller QFI than the MZI for all values of $\eta$.}
\label{EqFig}
\end{figure}

Experimentally, it is often the case that both nonlinear elements have the same squeezing magnitude since the same pump is used for both. In this section we investigate what the optimal QFI is when we restrict $r_1=r_2\equiv r$. 

The Yurke scheme is optimized when $\varphi=\pi$, resulting in a QFI that depends on the external loss quadratically:
\begin{equation}
    I_Q=\eta\sinh^2(2r)(2-\eta).
\end{equation}
This is largest when we have no seeding, giving an optimal value of $I_Q=4\eta(2-\eta)N_\phi(N_\phi+1)=\eta(2-\eta)I_{Q_{\mathrm{max}}}$. We see in Fig.~\ref{EqFig} that the Yurke scheme again reigns supreme over the Mandel scheme. If mode $\hat{a}$ is included, the Mandel scheme surpasses the MZI only when $\eta$ is above some critical value (which decreases when $N_\phi$ is larger, as shown in Appendix \ref{app:supp fig}); without mode $\hat{a}$ the Mandel scheme falls short of the MZI.

\section{Detection Scheme for the Mandel Interferometer}
\label{sec: detection scheme}

A specific measurement scheme for the light exiting an interferometer will not necessarily achieve the minium uncertainty given by the QFI. Moreover, the parameters that optimize the QFI, such as $\varphi$ and $r_2$, are not necessarily the easiest to experimentally implement. There are thus additional considerations for each interferometric setup that are not all contained in the QFI. 

\begin{figure*}[hbt!]
    \includegraphics[width=\textwidth,trim={1cm 1.4cm 0 0},clip]{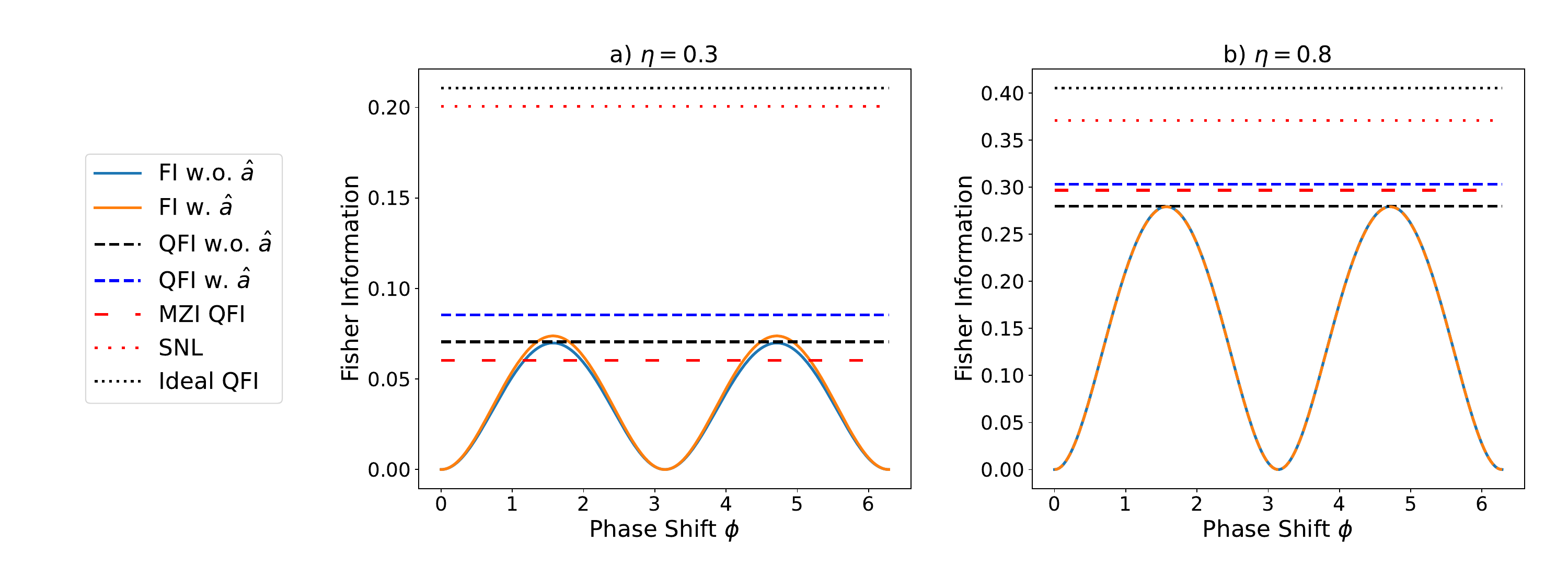}    
\caption{Fisher information vs. phase shift $\phi$ of the Mandel interferometer with external loss detected with photon number resolving detectors (PNRDs). We show the curve when mode $\hat{a}$ is discarded (light blue, solid) and when mode $\hat{a}$ is detected (orange, solid). We compare with the QFI when mode $\hat{a}$ is discarded (black, dashed), the QFI when mode $\hat{a}$ is retained (blue, dashed), the QFI of the MZI (red, loosely dashed), the SNL (red, loosely dotted), and the ideal QFI at $\eta=1$ (black, dotted). When calculating the FI numerically we must set a maximum number that the PNRDs can detect, chosen here to be 15. (a) FI at $\eta=0.3$, $r_1=0.1$, $r_2=1.5$, $\alpha=0$, $\beta=2$, and $\gamma=0$. Unlike with the QFI where it was optimal to have $r_1=0$, a nonzero value of $r_2$ is helpful. When mode $\hat{a}$ is left undetected the FI can surpass the QFI of the MZI and saturate the QFI for certain values of $\phi$. Detecting mode $\hat{a}$ allows for an increase in the maximum value of the FI, but the QFI when mode $\hat{a}$ is retained is not saturated. (b) FI at $\eta=0.8$, $r_1=0.3$, $r_2=1$, $\alpha=0$, $\beta=0$, and $\gamma=0$. The FI can saturate the QFI when mode $\hat{a}$ is discarded. The FI when mode $\hat{a}$ is detected has an identical curve, suggesting that PNRDs are not capable of extracting the additional information contained in mode $\hat{a}$ when $\eta>\frac{1}{2}$.}
\label{CFIFig}
\end{figure*}

In this section, we consider detecting modes of the Mandel interferometer with photon number resolving detectors (PNRDs) by looking at the (classical) Fisher information. PNRDs are modern photodetectors that are crucially able to sense correlations between photon numbers and uncover quantum properties that Gaussian detectors cannot \cite{BenchmarkPNRD} (recall that the optimal measurement of a lossy Gaussian state is not necessarily Gaussian \cite{Ohetal2019}), and can be used in conjunction with squeezed vacuum states to surpass classical sensitivity limits in interferometry \cite{Fabry-Parot_PNRD}. The Fisher information, in turn, quantifies the average amount of information about our parameter $\phi$ in a specific measurement (in contrast to the QFI, which depends only on the quantum state) \cite{Fisher}. For a Fisher information $I_C(\phi)$, the uncertainty of $\phi$ obeys the Cramér–Rao bound
\begin{equation}
\label{Classical CRB}
    (\Delta\phi)^2\geq\frac{1}{I_C(\phi)}\geq\frac{1}{I_Q}.
\end{equation}

Using the Python libraries StrawberryFields \cite{strawberryfields} and The Walrus \cite{Walrus}, we numerically calculated the Fisher information of the Mandel scheme both when modes $\hat{b}$ and $\hat{c}$ are detected with PNRDs and when all modes are detected with PNRDs, as seen in Fig. ~\ref{CFIFig}. Regardless of whether mode $\hat{a}$ is detected, the Fisher information can surpass the QFI of the MZI for $\eta<\frac{1}{2}$ when $r_2$ is sufficiently large, which aligns with our analysis of the QFI. However, when $\eta>\frac{1}{2}$, we did not find any parameters for which the Fisher information surpasses the QFI of the MZI, even when the QFI of the Mandel scheme does. It appears that detection with PNRDs is sufficient for the Fisher information to saturate the QFI only when mode $\hat{a}$ is discarded.

\section{Conclusion}

We found analytical expressions for the quantum Fisher information (QFI) of the Yurke and Mandel interferometers with vacuum or coherent-state seeding. When there is no loss, the Yurke and Mandel interferometers have equivalent QFIs, which are not improved by coherent light seeding. They also have equivalent QFIs when there is internal loss, but seeding can improve the QFI in this case. A quantum advantage only exists above a certain threshold of internal loss and we quantified the dependence of this threshold on the average photon number illuminating the sample. The QFI of the Yurke interferometer is completely robust to external loss provided that the second squeezing magnitude $r_2$ can be made large enough.  The Mandel interferometer is robust to external loss above $50\%$ if $r_2$ is large when mode $\hat{a}$ in included; if mode $\hat{a}$ is discarded then it is still robust but only after a higher threshold of loss (which depends on $r_2$). In all three cases, the QFI is robust to loss up to order $\frac{e^{-2r_2}}{\eta}$. When external losses are less than $50\%$, the Mandel interferometer can achieve a quantum advantage if mode $\hat{a}$ is included, otherwise it has equivalent QFI to the  Mach-Zehnder Interferometer.  When modes $\hat{b}$ and $\hat{c}$ of the Mandel interferometer are detected with photon number resolving detectors the Fisher information can saturate the QFI; we did not find any parameters where the QFI is saturated if mode $\hat{a}$ is detected.

We here considered only the dominant source of noise for interferometry, loss, which encompasses absorption, mode mismatch, detector inefficiencies, and more. Further work could study the effects of seeding in nonlinear interferometry with phase noise or decoherence, even though  experts at building interferometers tend to be very good at avoiding these types of noise. Whereas here seeding helps overcome loss by directly adding more photons, in that scenario, it would be interesting to investigate whether seeding could ``re-inject'' some coherence into a system that has lost some. As for other imperfections such as detector dark counts, it remains to be seen whether such noise can be surmounted by seeded nonlinear interferometry.

Along the way, we developed code that efficiently computes symbolic expressions for the QFI of Gaussian interferometers and is now publicly available at \url{https://github.com/JKranias/QFI-of-Gaussian-Interferometers.git}. The expressions can thence be optimized with respect to arbitrary parameters, including arbitrary functions of $N_\phi$, loss parameters, achievable nonlinearities, and more. We thus expect our work to find application both in the theoretical studies of what makes an interferometer better than another and practical calculations for relevant phase-sensing experiments.

\section*{Acknowledgments}

The authors acknowledge that the NRC headquarters is located on the traditional unceded territory of the Algonquin Anishinaabe and Mohawk people. This work was supported by the NRC's Quantum Sensors challenge program and the NRC's student employment program. The authors thank Barry Sanders and Myungshik Kim for discussions that led to initiating this work and Raj Patel, Gerard Jimenez Machado, and Anaelle Hertz for fruitful discussions. AZG acknowledges funding from the NSERC PDF program. KH acknowledges funding from NSERC’s Discovery Grant.

\bibliography{bibliography}

\onecolumngrid

\begin{appendix}

\section{Useful Identities}
In deriving the QFI for Gaussian states, we rely on a number of properties. For the symplectic matrix $K=\begin{pmatrix}
    \mathds{I}_n &0\\ 0&-\mathds{I}_n
\end{pmatrix}$, we have
\begin{equation}
\label{Kinv}
    K^{-1}=K^T=\overline{K}=K.
\end{equation}
Then, for all matrices $A$, $B$, and $C$, we have:
\begin{equation}
\label{kroninv}
    (A\otimes B)^{-1}=(A^{-1}\otimes B^{-1}),
\end{equation}
\begin{equation}
\label{kronprod}
    (AB)\otimes(CD)=(A\otimes C)(B\otimes D),
\end{equation}
\begin{equation}
\label{veckron}
    (C^T\otimes A)\text{vec}\!\left[  B  \right]=\text{vec}\!\left[  ABC  \right], \;\;\; \text{vec}\!\left[  B  \right]^\dagger(\overline{C}\otimes A^\dagger)=\text{vec}\!\left[  ABC  \right]^\dagger,
\end{equation}
\begin{equation}
\label{vecconj}
    \text{vec}\!\left[  A  \right]^\dagger=\text{vec}\!\left[  \overline{A}  \right]^T.
\end{equation}

\section{Lossy QFI Alternate Form Derivation} \label{AA}
By combining Eqs.~\eqref{Kinv} and \eqref{kroninv}, we know that $(K\otimes K)^{-1}=(K\otimes K)$. This can be used to rewrite $\mathcal{M}^{-1}$:
\begin{equation}
    %\begin{align*}
    \mathcal{M}^{-1}=(\overline{\sigma}\otimes\sigma-K\otimes K)^{-1}
%\\
=(((\overline{\sigma}K\otimes\sigma K)-\mathbb{I})(K\otimes K))^{-1}
%\\
=(K\otimes K)((\overline{\sigma}K\otimes\sigma K)-\mathbb{I})^{-1}.
%\end{align*}
\end{equation}
If we have the eigendecomposition $(\overline{\sigma}K\otimes\sigma K)=Q\Lambda Q^{-1}$, then
\begin{equation}
    %\begin{align*}
    \mathcal{M}^{-1}=(K\otimes K)(Q\Lambda Q^{-1}-\mathbb{I})^{-1}
    %\\
    =(K\otimes K)(Q(\Lambda-\mathbb{I})Q^{-1})^{-1}
    %\\
    =(K\otimes K)Q(\Lambda-\mathbb{I})^{-1}Q^{-1}.
%\end{align*}
\end{equation}
Consider the eigendecomposition $\sigma K=q\lambda q^{-1}$, and note that the symplectic eigenvalues must be real, therefore $\lambda$ is real. Also note $\overline{\sigma}=\sigma^T$ since $\sigma$ is Hermitian. Now 
\begin{equation}
%    \begin{align}
\label{sigmadecomp}
    \sigma=\sigma KK %\\
    =q\lambda q ^{-1}K, 
%\end{align}
\end{equation}
therefore 
\begin{align}
    \overline{\sigma}K=\sigma^TK%\\
    =(q\lambda q^{-1}K)^TK%\\
    =K(q^{-1})^T\lambda q^TK.
\end{align}
This implies
\begin{align}
    (\overline{\sigma}K\otimes\sigma K)=(K(q^{-1})^T\lambda q^TK)\otimes(q\lambda q^{-1})%\\
    =(K(q^{-1})^T\otimes q)(\lambda\otimes \lambda)(q^TK\otimes q^{-1})%\\
    =Q\Lambda Q^{-1},
\end{align}
where $Q=(K(q^{-1})^T\otimes q)$ and $\Lambda=\lambda\otimes \lambda$.
Note that $\lambda\otimes \lambda$ is diagonal, therefore we have the eigendecomposition of $(\overline{\sigma}K\otimes\sigma K)$.

Now $(K\otimes K)Q=(K\otimes K)(K(q^{-1})^T\otimes q)=((q^{-1})^T\otimes Kq)$. We can therefore write:
\begin{equation}
    \begin{aligned}
    \text{vec}\!\left[  \partial_\phi\sigma  \right]^\dagger \mathcal{M}^{-1}\text{vec}\!\left[  \partial_\phi\sigma  \right]
    %\\
    &=\text{vec}\!\left[  \partial_\phi\sigma  \right]^\dagger((q^{-1})^T\otimes Kq)(\lambda\otimes\lambda-\mathbb{I})^{-1}(q^TK\otimes q^{-1})\text{vec}\!\left[  \partial_\phi\sigma  \right]
    \\
    &=\text{vec}\!\left[  q^TK(\partial_\phi\sigma)^T(q^{-1})^T  \right]^T(\lambda\otimes\lambda-\mathbb{I})^{-1}\text{vec}\!\left[  q^{-1}(\partial_\phi\sigma)Kq  \right]
\end{aligned}
\end{equation}
where we used Eqs.~\eqref{veckron} and \eqref{vecconj}.

Now let $\Sigma=q^{-1}(\partial_\phi\sigma)Kq$, and we can write the QFI as
\begin{equation}
\label{altQFIv1}
    I_Q=\frac{1}{2}\text{vec}\!\left[  \Sigma^T  \right]^T(\lambda\otimes\lambda-\mathbb{I})^{-1}\text{vec}\!\left[  \Sigma  \right]+2(\partial_\phi\textbf{d})^\dagger\sigma^{-1}(\partial_\phi\textbf{d}).
\end{equation}
This is similar to Eq.~(E2) in \cite{Safranek} but we used the eigendecomposition of $\sigma K$ instead of the symplectic decomposition of $\sigma$. The advantage of using the eigendecomposition is that we do not need to compute $\sigma^{\frac{1}{2}}$.

Since the displacement vector has the structure $\boldsymbol{d}=\begin{pmatrix}\pmb{\gamma} \\ \overline{\pmb{\gamma}}\end{pmatrix}$ \cite{Safranek}, its Hermitian conjugate is given by the rearrangement
\begin{align}
(\partial_\phi\boldsymbol{d})^\dagger=\begin{pmatrix}\partial_\phi \overline{\pmb{\gamma}}&\partial_\phi \pmb{\gamma}\end{pmatrix}=(\partial_\phi\boldsymbol{d})^T\begin{pmatrix}0&\mathbb{I}_n\\ \mathbb{I}_n&0 \end{pmatrix}. 
\end{align}
And using Eq.~\eqref{sigmadecomp}, the second term of the QFI can be written as
\begin{align}
     2(\partial_\phi\textbf{d})^\dagger\sigma^{-1}(\partial_\phi\textbf{d})=2(\partial_\phi\boldsymbol{d} )^T\begin{pmatrix}0&\mathbb{I}_n\\ \mathbb{I}_n&0 \end{pmatrix}Kq\lambda^{-1}q^{-1}(\partial_\phi\textbf{d})\\
     =2(\partial_\phi\boldsymbol{d} )^T\begin{pmatrix}0&-\mathbb{I}_n\\ \mathbb{I}_n&0 \end{pmatrix}q\lambda^{-1}q^{-1}(\partial_\phi\textbf{d}).
\end{align}
Combining this with Eq.~\eqref{altQFIv1} we get Eq.~\eqref{altQFI}. Although omitting the use of the Hermitian conjugate makes the expression less elegant, it is more convenient to evaluate in practice since symbolic algebra programs such as Mathematica do not always fully simplify the complex conjugate of symbolic expressions.

\section{Lossless QFI Calculation}\label{AB}
A Mathematica notebook was used to find the output Gaussian state as an analytic expression depending on all the parameters (seeding, squeezing, and loss). Applying this to Eq.~\eqref{PureGaussQFI}, we find \begin{equation}
    I_Q=4\cosh ^4r_1|\alpha|^2+4\sinh ^4r_1|\beta|^2+\sinh ^2(2r_1)(|\alpha|^2+|\beta|^2+1)-8\cosh r_1\sinh r_1(\cosh ^2r_1+\sinh ^2r_1)\RE(\alpha\beta).
\end{equation}
The average intensity of light that passes through the sample in mode a is
%\begin{widetext}
\begin{equation}
\label{AppendixNphi}
    N_\phi=\bra{\alpha,\beta}S_{ab}^\dagger(r_1) a^\dagger aS_{ab}(r_1)\ket{\alpha,\beta}=\cosh ^2r_1|\alpha|^2+\sinh ^2r_1(|\beta|^2+1)-2\cosh r_1\sinh r_1\RE(\alpha\beta),
\end{equation}
%\end{widetext}
where $S_{ab}(r_1)=e^{r_1(ab-a^\dagger b^\dagger)}$ is the two-mode squeezing operator. By rearranging, we find that 
\begin{equation}
\label{appendixlosslessIQ}
    I_Q=4\cosh ^2r_1N_\phi+4\sinh ^2r_1N_\phi-4\sinh ^4r_1.
\end{equation}

If we keep $N_\phi$ fixed, then $I_Q$ is maximized when $\sinh ^2r_1=N_\phi$. Therefore the maximum QFI is
\begin{equation}
    I_{Q_{\mathrm{max}}}=4\cosh ^2r_1\sinh ^2r_1=4N_\phi(N_\phi+1).
\end{equation}
Achieving $\sinh ^2r_1=N_\phi$ is simplest when $\alpha=\beta=0$, but Eq.~\eqref{AppendixNphi} can also be satisfied for nonzero seeding under the constraint $\sinh ^2r_1=N_\phi$, as long as $\cosh^2r_1|\alpha|^2+\sinh ^2r_1|\beta|^2-2\cosh r_1\sinh r_1\RE(\alpha\beta)=|\alpha \cosh r_1-\beta^* \sinh r_1|^2=0$. And since $\sinh r_1=\sqrt{N_\phi}$, $\cosh r_1=\sqrt{N_\phi+1}$, the condition for optimal seeding is
\begin{equation}
    \alpha=\sqrt{\frac{N_\phi}{N_\phi+1}}\beta^*.
\end{equation}
Also note that for fixed $N_\phi$, $r_1$ is bounded by $\sinh ^2r_{1_{\mathrm{max}}}=N_\phi$. Eq.~\eqref{appendixlosslessIQ} is increasing for all $r_1$ in $\!\left[ 0,r_{1_{\mathrm{max}}} \right]$, and when $r_1=0$ the QFI is $I_Q=4N_\phi$, which is the same as the QFI of the MZI. Therefore, there is a quantum advantage in the QFI for any $r_1>0$, and making $r_1$ as large as possible will result in the best QFI.

\section{Lossy QFI Calculation}\label{AC}
The Gaussian output state of the interferometer is found using the methods outlined in \ref{GS}. The output state can then be used to find the QFI from Eq.~\eqref{altQFI} or Eq.~\eqref{2modeQFI}.

In all the cases we considered, the QFI's dependence on seeding could be directly related to $N_\phi$, allowing $I_Q$ to be written in terms of only $N_\phi$, $\phi$, and the squeezing parameters. Since no loss occurs before the phase shift, Eq.~\eqref{Nphi} remains valid. Therefore, the highest $r_1$ possible for a given $N_\phi$ is
\begin{equation}
    r_{1_{\mathrm{max}}}=\text{arcsinh} (\sqrt{N_\phi}),
\end{equation}
which occurs when $\alpha=\beta=0$. Therefore, $I_Q$ is optimized under the constraint $0\leq r_1\leq \text{arcsinh} (\sqrt{N_\phi})$ for a fixed $N_\phi$.

When $I_Q$ is optimized at $r_{1\mathrm{opt}} \neq r_{1_{\mathrm{max}}}$, then there must be seeding to satisfy Eq.~\eqref{Nphi}. Eq.~\eqref{Nphi} can be satisfied for $\beta=0$, $|\alpha|^2=\frac{N_\phi-\sinh ^2r_{1\mathrm{opt}} }{\cosh ^2r_{1\mathrm{opt}} }$. This means that the optimal QFI can be achieved by only seeding mode $\hat{a}$, and only the magnitude of the seeding is relevant, not the phase. Note that it is still possible to satisfy Eq.~\eqref{Nphi} and achieve optimal seeding while seeding both modes, or only mode $\hat{b}$ (when $r_{1\mathrm{opt}} \neq0$), but only seeding mode $\hat{a}$ is the most simple method. Additionally, large values of $|\beta|^2$ are necessary to satisfy Eq.~\eqref{Nphi} when $r_{1\mathrm{opt}} $ is small if mode $\hat{a}$ is not seeded.

\section{Calculations for Yurke and Mandel Interferometers with Internal Loss}\label{AD}

The QFI (Eq.~\eqref{IQA}) was found using Eq.~\eqref{altQFI}. 

If $N_\phi\leq\frac{1-T}{2T(2T-1)}$, then the QFI is optimized at $r=0$, and we have the maxiumum QFI as $I_Q=4TN_\phi$, the same as for the lossy MZI. (This condition was found by evaluating the second derivative at $r=0$.) Rearranging the inequality, we have the critical $T$ to surpass the MZI as 
\begin{equation}
    T_C=\frac{(2N_\phi-1)}{8N_\phi}\left(1\pm\sqrt{1+\frac{16N_\phi}{(2N_\phi-1)^2}}\right),
\end{equation} 
where we take the $+(-)$ case for $N_\phi$ above (below) $N_\phi=0.5$. Note that $T_C\rightarrow0.5$ as $N_\phi\rightarrow\infty$, therefore a quantum advantage can never be gained for $T\leq0.5$.

\section{Calculations for Yurke Interferometer with External Loss}\label{AE}

The QFI (Eq.~\eqref{EY}) was found using Eq.~\eqref{2modeQFI}. 

Plotting Eq.~\eqref{EY} numerically shows that it can  become arbitrarily close to $I_{Q_{\mathrm{max}}}$ at any nonzero value of $\eta$ when $r_2$ is sufficiently large and there is no seeding ($\sinh^2(r_1)=N_\phi$). If we set $\sinh^2(r_1)=N_\phi$ and take the derivative of the QFI with respect to $\cos\varphi$, then expand for $e^{2r_2}\gg1$, we find the optimal value in Eq.~\eqref{popt}:
\begin{equation}
    \cos\varphi_0=\frac{-2\sqrt{N_\phi(N_\phi+1)}}{2N_\phi+1}+\frac{(1-\eta)\sqrt{N_\phi(N_\phi+1)}}{2(2N_\phi+1)^2\eta\cosh^2(r_2)}+\mathcal{O}\bigg(\frac{e^{-4r_2}}{\eta^2}\bigg).
\end{equation}
Inserting this back into the QFI gives
\begin{equation}
    I_Q=\sinh^2(2r_1)\left(1-\frac{(1-\eta)(2N_\phi+1)}{2\eta\cosh^2(r_2)}+\mathcal{O}\bigg(\frac{e^{-4r_2}}{\eta^2}\bigg)\right).
\end{equation}

\section{Calculations for Mandel Interferometer with External Loss, Mode $\hat{a}$ Discarded}

The QFI (Eq.~\eqref{MEnoa}) was found using Eq.~\eqref{2modeQFI}. Note that this does not have to reduce to the lossless case (Eq.~\eqref{LosslessQFI}) when $\eta=1$ since we do not assume access to all modes.

Since $\sinh(r_2)<\cosh(r_2)$, and  $\sinh(r_2)$ gets closer to $\cosh(r_2)$ monotonically as $r_2$ becomes larger, both terms of Eq.~\eqref{MEnoa} will be largest when  $r_2$ is as large as possible. Therefore to maximize the QFI, we take $r_2$ sufficiently large such that $\sinh(r_2)\approx\cosh(r_2)$ (i.e. $e^{2r_2}\gg1$). The derivative with respect to $r_1$ is negative for $\eta<\eta_0$ and positive for $\eta>\eta_0$, where
\begin{equation}
    \eta_0=\frac{\Omega+\sqrt{\Omega^2+16(N_\phi+1)\cosh^2(r_1)\cosh^2(r_2)(\cosh^2(r_1)\cosh^2(r_2)-1)^2(\cosh^2(r_1)\cosh^2(r_2)-2)}}{8(N_\phi+1)\cosh^2(r_2)(\cosh^2(r_1)\cosh^2(r_2)-1)^2},
\end{equation}
and $\Omega=2(N_\phi+1)\cosh^2(r_2)(\cosh^2(r_1)\cosh^2(r_2)-1)^2-4\cosh^4(r_1)\cosh^4(r_2)+4\cosh^2(r_1)\cosh^2(r_2)$. When $e^{2r_2}\gg1$, $\eta_0$ reduces to
\begin{equation}
    \eta_0=\frac{1}{2}-\frac{1}{2(N_\phi+1)\cosh^2(r_2)}+\mathcal{O}(e^{-4r_2}).
\end{equation}
This is independent of $r_1$ up to order $e^{-4r_2}$, Therefore, except for within a small range near $\eta_0$, when $\eta>\eta_0$ the optimal QFI is achieved at $r_1=r_{1_{\mathrm{max}}}=\text{arcsinh}(\sqrt{N_\phi})$ with the value 
\begin{equation}
    I_Q=\frac{4\eta N_\phi(N_\phi+1)\sinh^2(r_2)}{(N_\phi+1)\cosh^2(r_2)-1},
\end{equation}
which is equal to
\begin{equation}
    I_Q=4\eta N_\phi\bigg(1-\frac{N_\phi}{(N_\phi+1)\cosh^2(r_2)}+\mathcal{O}(e^{-4r_2})\bigg)
\end{equation}
when $\cosh^2(r_2)\gg1$, identical to the QFI of the MZI up to order $e^{-2r_2}$. . However, in the case $\eta<\eta_0$ (again exclding a small range near $\eta_0$), the optimal QFI is at $r_1=0$ and has the value 
\begin{equation}
    I_Q=\frac{4\eta N_\phi\sinh^2(r_2)}{1+2\eta\sinh^2(r_2)},
\end{equation}
which approaches
\begin{equation}
    I_Q=2N_\phi\left(1-\frac{1}{2\eta\sinh^2(r_2)}+\mathcal{O}\bigg(\frac{e^{-4r_2}}{\eta^2}\bigg)\right)
\end{equation}
when $\sinh^2(r_2)\gg\frac{1}{\eta}$. We see the small range where $\eta_0$ is dependant on $r_1$ manifest in our numerical optimization with a small range near $\eta_0$ where the optimal $r_1$ value varies continuously from $0$ to $r_{1_{\mathrm{max}}}$. The Mandel scheme without mode $\hat{a}$ has an advantage over the MZI only for $\eta<\eta_0$.

\section{Calculations for Mandel Interferometer with External Loss, All Modes Retained}
The QFI was found using Eq.~\eqref{altQFI}. From numerical optimization of Eq.~\eqref{MQFI} it is clear that if the maximum allowed value of $r_2$ is sufficiently large (which occurs beyond $e^{2r_2}\gg1$), then the optimal value of $r_1$ is $r_1=0$ when $\eta<\frac{1}{2}$, and $r_1=r_{1_{\mathrm{max}}}=\text{arcsinh}(\sqrt{N_\phi})$ when $\eta>\frac{1}{2}$.

If $r_1=0$, then
\begin{equation}
    I_Q=\frac{4N_\phi\eta(1+2(1-\eta)\sinh^2(r_2))}{1+4\eta(1-\eta)\sinh^2(r_2)},
\end{equation}
and
\begin{equation}
    \partial_{r_2}I_Q=\frac{8\eta(1-\eta)(1-2\eta)N_\phi\sinh(2r_2)}{(1+4\eta(1-\eta)\sinh^2(r_2))^2},
\end{equation}
which is always positive for $\eta<\frac{1}{2}$, therefore the maximum QFI for $\eta<\frac{1}{2}$ is achieved when $r_2$ is as large as possible. When $e^{2r_2}\gg\frac{1}{\eta(1-\eta)}$ the QFI reduces to
\begin{equation}
    I_Q=2N_\phi\left(1-\frac{(1-2\eta)}{4\eta(1-\eta)\sinh^2(r_2)}+\mathcal{O}\bigg(\frac{e^{-4r_2}}{\eta^2}\bigg)\right).
\end{equation}
If $r_1=r_{1_{\mathrm{max}}}$, then the solutions for $r_2$ of $\partial_{r_2}I_Q=0$ are
\begin{equation}
    r_2^{(0)}=0,\;\;r_2^{(1)}=\text{arccosh}\left(\sqrt{\frac{2(1-2\eta(1-\eta))+\sqrt{2(1-2\eta(1-\eta))}}{2\eta(2\eta-1)(1+N_\phi)}}\right).
\end{equation}
When $r_1=r_{1_{\mathrm{max}}}$ and $r_2=r_2^{(0)}$, then
%\begin{widetext}
%
\begin{equation}
    I_Q=\frac{4\eta^2 N_\phi(N_\phi+1)}{1+2N_\phi\eta(1-\eta)}.
\end{equation}
%\end{widetext}
When $r_1=r_{1_{\mathrm{max}}}$ and $r_2=r_2^{(1)}$,
%\begin{widetext}
%
\begin{equation}
    I_Q=4\eta N_\phi\left(1+\eta N_\phi\left(1-2(1-\eta)(2\eta-1)-2(1-\eta)\sqrt{2-4\eta(1-\eta)}\right)\right).
\end{equation}
%\end{widetext}
which is always greater than the $r_2^{(0)}$ solution. However, if 
\begin{equation}
    \frac{2(1-2\eta(1-\eta))+\sqrt{2(1-2\eta(1-\eta))}}{2\eta(2\eta-1)(1+N_\phi)}<1 \implies N_\phi>\frac{2(1-2\eta(1-\eta))+\sqrt{2(1-2\eta(1-\eta))}}{2\eta(2\eta-1)}-1
\end{equation}
then $r_2^{(1)}$ is not a valid solution since $\text{arcosh}(x)$ is undefined for $x<1$. This is how we got the maximum QFI and the condition Eq.~\eqref{r20cond}.

\newpage
\section{Supplemental Figures}
\label{app:supp fig}
\begin{figure*}[htb]
\centering
%viewport =0 45 600 210
\includegraphics*[width=\textwidth,trim={1cm 1.5cm 0 0},clip]{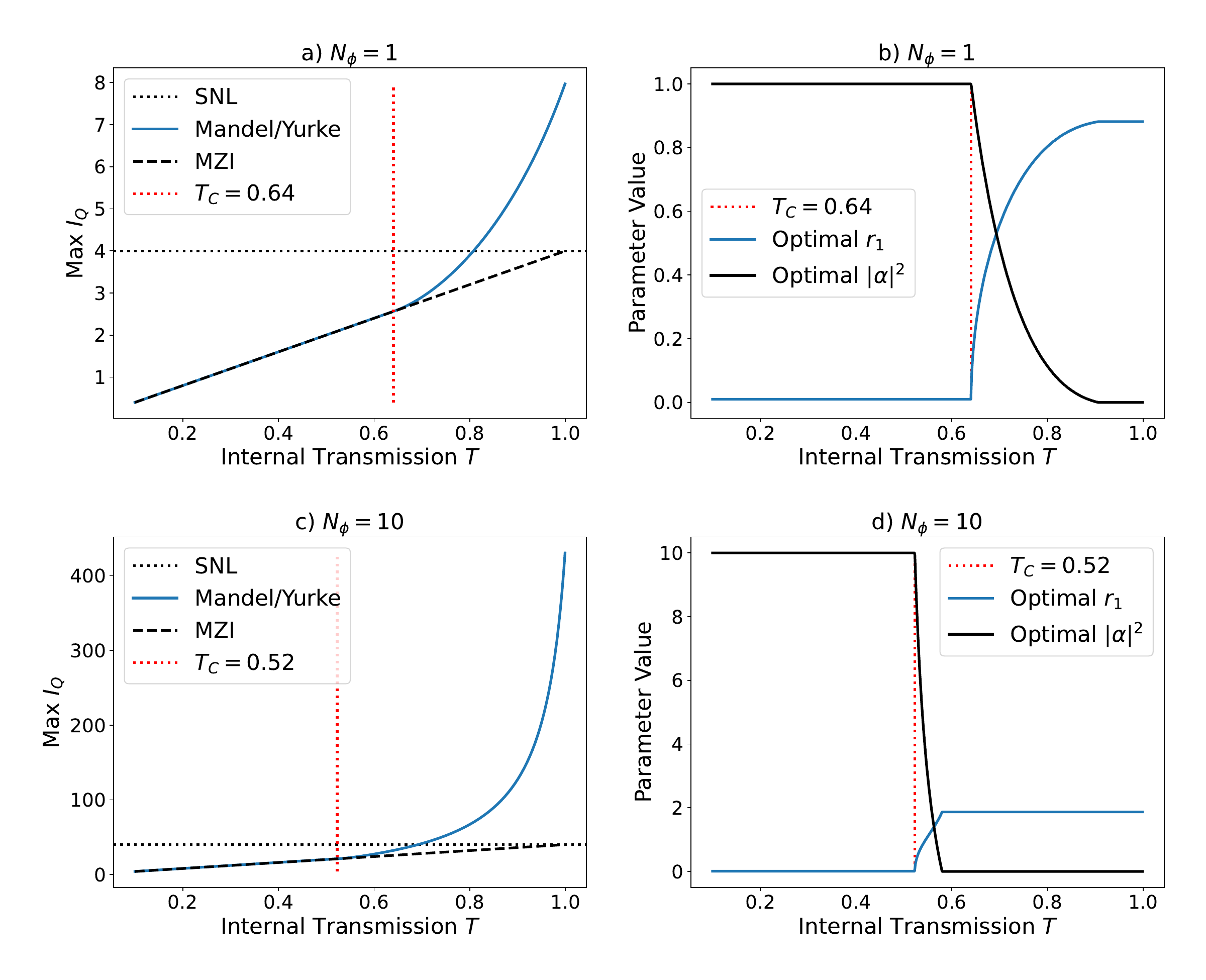}
  \caption{(a) Recreation of Fig.~\ref{InternalFig}(a) when $N_\phi=1$. (b) Recreation of Fig.~\ref{InternalFig}(b) when $N_\phi=1$. (c) Recreation of Fig.~\ref{InternalFig}(a) when $N_\phi=10$. (d) Recreation of Fig.~\ref{InternalFig}(b) when $N_\phi=10$. Observe that $T_C$ decreases as $N_\phi$ increases, and is close to $\frac{1}{2}$ when $N_\phi=10$. Moreover, the QFI surpasses the SNL at lower values of $T$ as $N_\phi$ increases.}
  \label{SuppInternalFig}
  \end{figure*}

\begin{figure*}[htb]
\centering
%viewport =0 45 600 210
\includegraphics*[width=\textwidth,trim={1cm 1.5cm 0 0},clip]{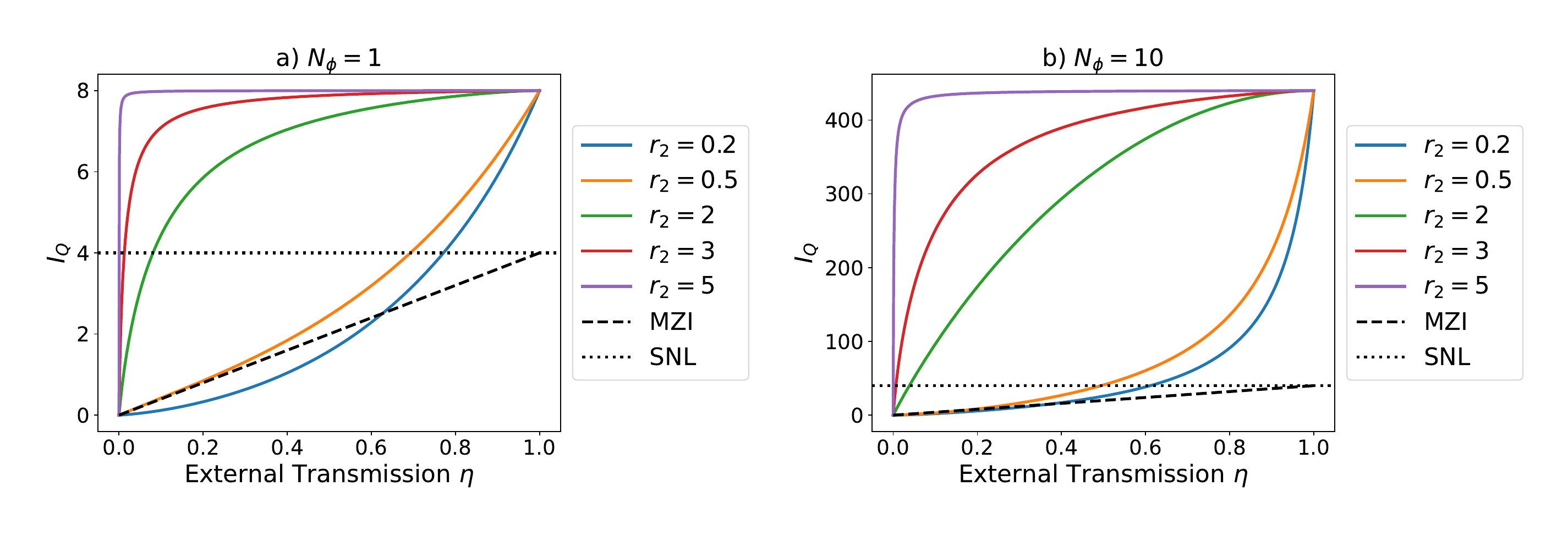}
  \caption{(a) Recreation of Fig.~\ref{YurkeExternalFig}(a) when $N_\phi=1$. (b) Recreation of Fig.~\ref{YurkeExternalFig}(a) when $N_\phi=10$. When $N_\phi$ increases the QFI for a given $r_2$ decreases relative to the maximum QFI, ie. $\frac{I_Q}{I_{Q_{\mathrm{max}}}}=\frac{I_Q}{4N_\phi(N_\phi+1)}$ decreases. ~\ref{AE}  However, the QFI for a given $r_2$ surpasses the SNL at a lower value of $\eta$ when $N_\phi$ is larger, similarly to the case of internal loss. Increasing $N_\phi$ can be beneficial or detrimental, depending on if the priority is surpassing the SNL, or coming close to $I_{Q_{\mathrm{max}}}$.}
  \label{SuppYurkeExternalFig}
  \end{figure*}

\begin{figure*}[htb]
\centering
%viewport =0 45 600 210
\includegraphics*[width=\textwidth,trim={1cm 1.5cm 0 0},clip]{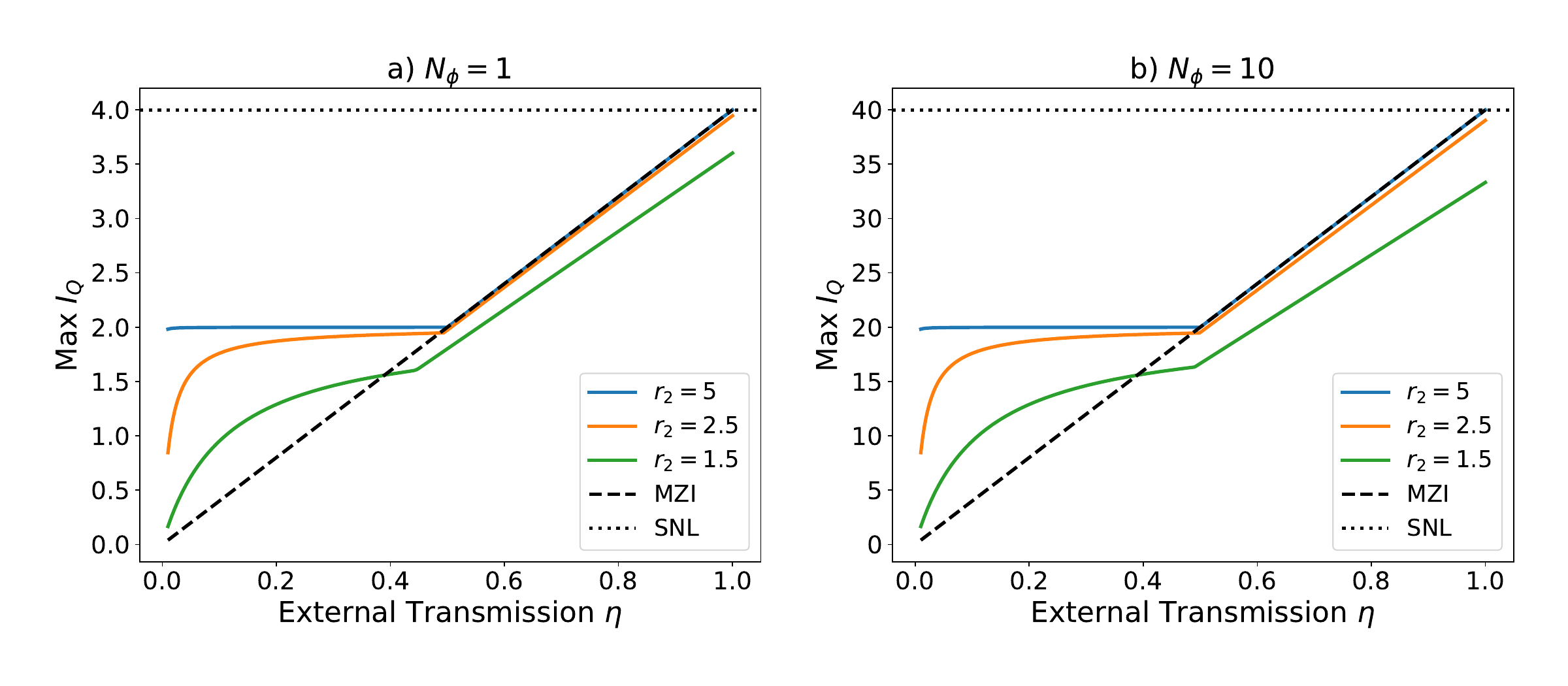}
  \caption{(a) Recreation of Fig.~\ref{MandelnoaFig} when $N_\phi=1$. (b) Recreation of Fig.~\ref{MandelnoaFig} when $N_\phi=10$. When $N_\phi$ is larger the QFI decreases more for smaller values of $r_2$, but $\eta_0$ stays closer to $\frac{1}{2}$.}
  \label{SuppMandelnoaFig}
  \end{figure*}

\begin{figure*}[htb]
\centering
%viewport =0 45 600 210
\includegraphics*[width=\textwidth,trim={1cm 1.5cm 0 0},clip]{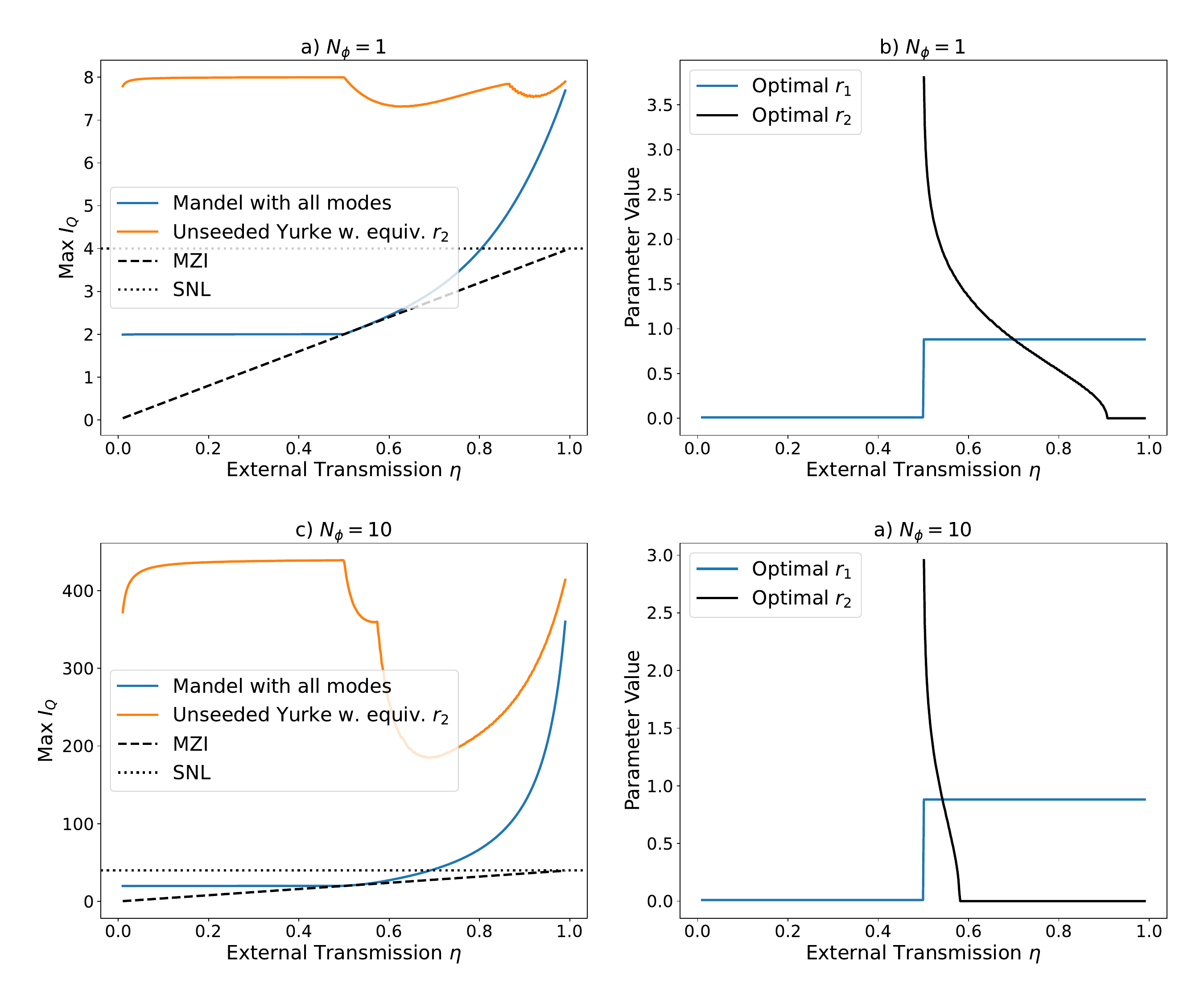}
  \caption{(a) Recreation of Fig.~\ref{MandelyesaFig}(a) when $N_\phi=1$. (b) Recreation of Fig.~\ref{MandelyesaFig}(b) when $N_\phi=1$. (c) Recreation of Fig.~\ref{MandelyesaFig}(a) when $N_\phi=10$. (d) Recreation of Fig.~\ref{MandelyesaFig}(b) when $N_\phi=10$. As with internal loss and the Yurke interferometer with external loss, the QFI surpasses the SNL for lower values of $\eta$ when $N_\phi$ is larger. The QFI of the Mandel interferometer relative to the Yurke interferometer with equivalent $r_2$ is smaller as $N_\phi$ increases. Unlike when $N_\phi=0.1$, it is optimal to have no squeezing in the second nonlinear element ($r_2=0$) above a certain $\eta$, and we see that Eq.~\eqref{r20cond} will be satisfied for a larger range of $\eta$ when $N_\phi$ is larger.}
  \label{SuppMandelyesaFig}
  \end{figure*}

\begin{figure*}[htb]
\centering
%viewport =0 45 600 210
\includegraphics*[width=\textwidth,trim={1cm 1.5cm 0 0},clip]{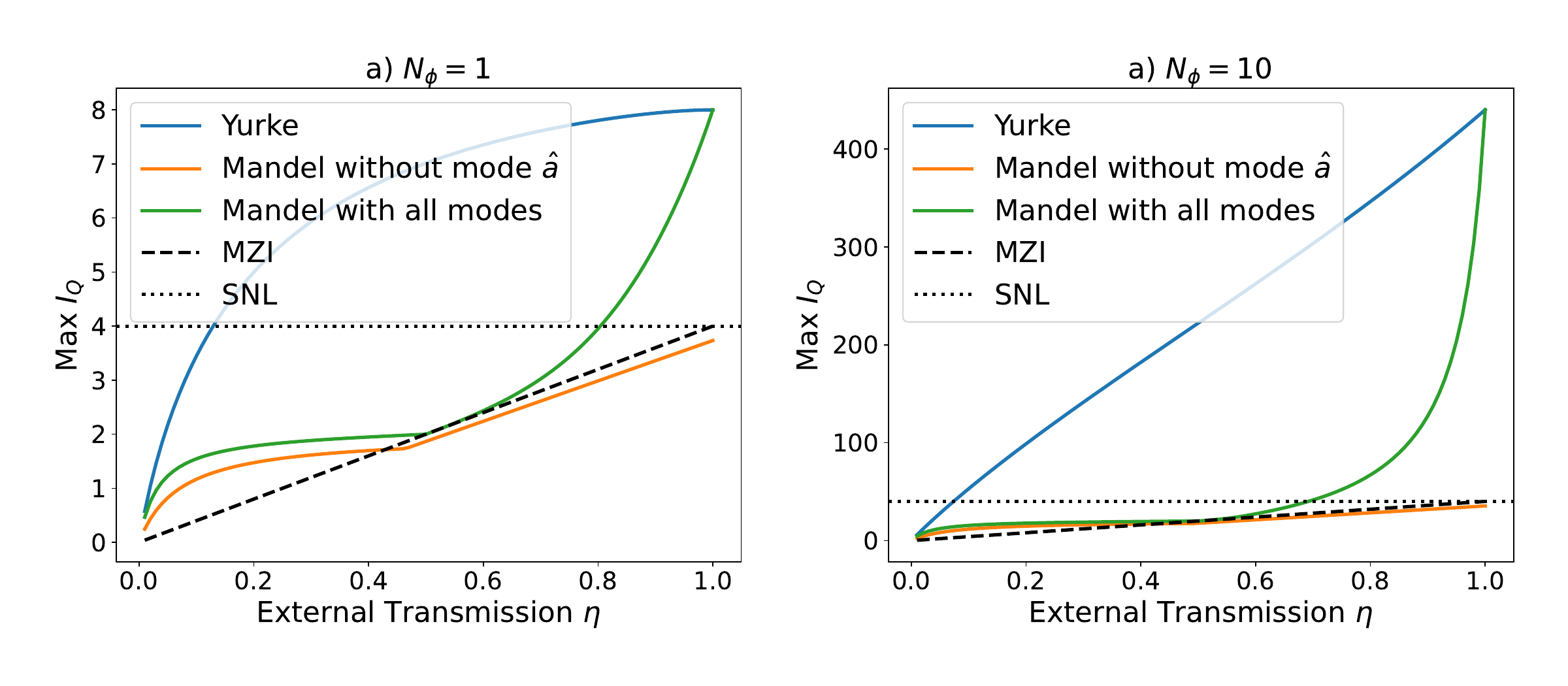}
  \caption{(a) Recreation of Fig.~\ref{ComparisonFig} when $N_\phi=1$. (b) Recreation of Fig.~\ref{ComparisonFig} when $N_\phi=10$. As in Fig.~\ref{SuppYurkeExternalFig} we see that the QFI of the Yurke scheme decreases relative to the maximum QFI as $N_\phi$ increases, but it still outperforms the Mandel scheme regardless of $N_\phi$.}
  \label{SuppComparisonFig}
  \end{figure*}

\begin{figure*}[htb]
\centering
%viewport =0 45 600 210
\includegraphics*[width=\textwidth,trim={1cm 1.5cm 0 0},clip]{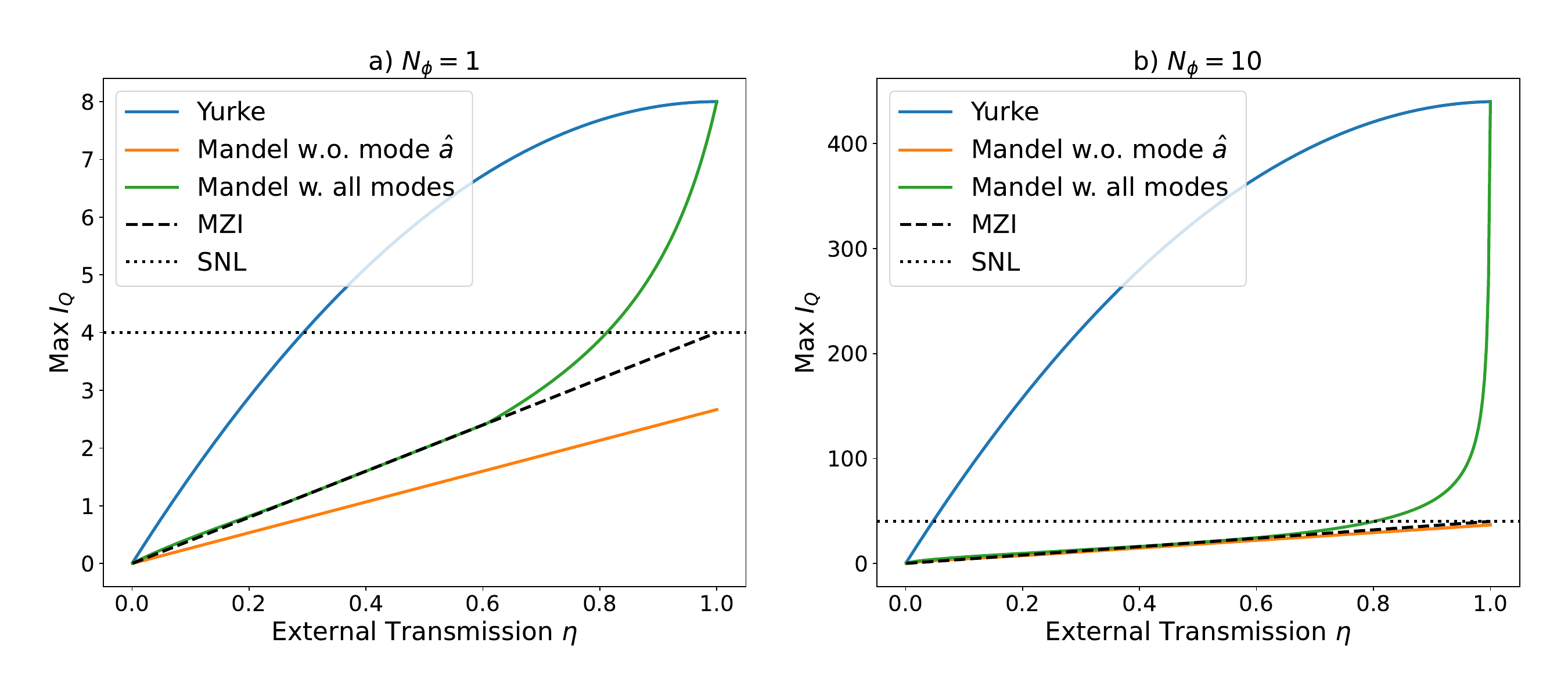}
  \caption{(a) Recreation of Fig.~\ref{EqFig} when $N_\phi=1$. (b) Recreation of Fig.~\ref{EqFig} when $N_\phi=10$. As $N_\phi$ increases, the Mandel scheme with all modes surpasses the MZI at smaller values of $\eta$.}
  \label{SuppEqFig}
  \end{figure*}

We here present figures similar to those in the main text but with different parameter values. Figures~\ref{SuppInternalFig}, \ref{SuppYurkeExternalFig}, \ref{SuppMandelnoaFig}, \ref{SuppMandelyesaFig}, \ref{SuppComparisonFig} and \ref{SuppEqFig} depict the same concepts as Figures~\ref{InternalFig}, ~\ref{YurkeExternalFig}, ~\ref{MandelnoaFig}, ~\ref{MandelyesaFig}, ~\ref{ComparisonFig} and ~\ref{EqFig}, respectively, with different values of $N_\phi$.

\end{appendix}

\end{document}